\documentclass[11.5pt]{article}
\usepackage{amsmath}
\usepackage{amsthm}
\usepackage{mathrsfs}
\usepackage{graphicx}
\usepackage{color}
\allowdisplaybreaks
\newtheorem{thm}{Theorem}[section] %(If you want theorem numbered
\newtheorem{lemma}{Lemma}[section] %%    with section number.
\newtheorem{cor}{Corollary}[section]

\newtheorem{definition}{Definition}[section]
\newcommand{\bed}{\begin{definition}}
\newcommand{\eed}{\end{definition}}

\newcommand{\bitem}{\begin{itemize}}
\newcommand{\eitem}{\end{itemize}}

\newcommand{\goto}{\rightarrow}

\newcommand{\beqn}{\begin{equation}}
\newcommand{\eeqn}{\end{equation}}
\newcommand{\balign}{\begin{align}}
\newcommand{\ealign}{\end{align}}

\usepackage{amssymb}
\newcommand{\beq}{\begin{equation}}
\newcommand{\eeq}{\end{equation}}

\newcommand{\diag}{\mathrm{diag}}

\usepackage{hyperref}
\hypersetup{
    colorlinks=true,
    linkcolor=blue,
    filecolor=magenta,      
    urlcolor=cyan,
}
\title{Improvements on SCORE, especially for weak signals}
\author{Jiashun Jin$^1$\and Zheng Tracy Ke$^2$ \and Shengming Luo$^1$}
\date{%
    {\small $^1$Carnegie Mellon University and 
    $^2$Harvard University}\\[2ex]%
    \today
}
\begin{document} 
%%%%%%%%%%%
\maketitle

\begin{abstract}
A network may have weak signals and  
severe degree heterogeneity, and may be very sparse in one occurrence but very dense in another. 
SCORE \cite{SCORE}  is a recent approach to network community detection. It accommodates severe degree heterogeneity and is  adaptive to different levels of sparsity, but its performance for networks with weak signals is unclear. 

In this paper, we show that in a broad class of network settings where we allow for weak signals, severe degree heterogeneity, and a wide range of network sparsity,   
SCORE achieves prefect clustering and has the so-called ``exponential rate" 
in Hamming clustering errors.  The proof uses the most recent advancement on 
entry-wise bounds for the leading eigenvectors of the network adjacency matrix. 

The theoretical analysis assures us that SCORE continues to 
work well in the weak signal settings, but it does not rule out the possibility 
that SCORE may be further improved to have better performance 
in real applications, especially for networks with weak signals.

As a second contribution of the paper, we propose 
SCORE+ as an improved version of SCORE.   
We investigate SCORE+ with $8$ network data sets and found 
that it outperforms several representative approaches.  In particular,   for the $6$ data sets with relatively strong signals, SCORE+ has similar performance as that of SCORE, but for the  $2$ data sets (Simmons, Caltech) with possibly weak signals,   SCORE+ has much lower error rates.  
SCORE+ proposes several changes to SCORE. We carefully explain the 
rationale underlying each of these changes, using a mixture of theoretical and numerical 
study. 
\end{abstract}

%%%%%%%%%%%%%%
%%%%%%%%%%%%%%
%%%%%%%%%%%%%%
\section{Introduction}
Community detection is a problem that has received considerable attention \cite{DCBM, JiZhuMM, bickel2009nonparametric}.  Consider an undirected network ${\cal N}$ and let $A$ be its adjacency matrix:  
\[
A(i,j) = \left\{ 
\begin{array}{ll} 
1, &\qquad \mbox{if there is an edge connecting  node $i$ and $j$},  \\
0, &\qquad \mbox{otherwise}.  
\end{array} 
\right. 
\]
Since the network is undirected,  $A$ is symmetrical. Also, as a convention, we do not consider self edges, so all diagonal entries of $A$ are $0$.  
We assume the network is connected, 
consisting of $K$ perceivable non-overlapping communities 
\[
{\cal C}_1, {\cal C}_2, \ldots, {\cal C}_K. 
\]
Following the convention  in many recent works on community detection (e.g., \cite{bickel2009nonparametric, JiZhuMM}), we assume $K$ as known and the nodes do not have mixed-memberships, so each node belongs to {\it exactly} one of 
the $K$ communities.   The community labels  are unknown, and the goal is to use $(A, K)$ to predict them.   In statistics, this is known as the clustering problem.

See \cite{jin2020estimating} and reference therein for discussions on how to estimate $K$, and \cite{MSCORE} for the generalization of SCORE for network analysis in the presence of mixed memberships. 
 
Similar to  ``cluster",   ``community" is a concept that is scientifically meaningful but mathematically hard to define. Intuitively, communities are clusters of nodes that have more edges ``within" than ``across" \cite{SCORE, zhao2012consistency}.  Note that ``communities"  and ``components"  are different concepts:  two  communities may be  connected,  while two  components are always disconnected. 

Table~\ref{tab:realdata0} presents $8$ network data sets which we analyze in this paper.  
Data sets 2-3  are from \cite{Traud2011,Traud2012} (see also \cite{chen2015convexified, ma2020universal}), and the other  $6$ datasets are downloaded from \url{http://www-personal.umich.edu/~mejn/netdata/}. 
For all these data sets, the true labels are suggested by the original authors or data curators, and we use the labels as the ``ground truth."  

%%%%%%%%%
%%%%%%%%%
%%%%%%%%% 
\begin{table}[htb!]  
\label{tab:realdata0} 
\centering
\scalebox{.95}{
\begin{tabular}{l|l|c|c|c|ccc}
Dataset & Source &  \#Nodes & $K$ &\#Edges & $d_{min}$  & $d_{max}$ & $\bar{d}$  \\
\hline
Weblogs &Adamic $\&$ Glance (2005) &  1222 & 2 & 16714  &   1 & 351 & 27.35\\
Simmons & Traud {\it et al}. (2011) &  1137 &4 &  24257 &    1&293&42.67   \\
Caltech &  Traud {\it et al}. (2011)  &  590 & 8 &   12822 &  1&179&43.36   \\
Football & Girvan \& Newman (2002) &  110  &11 & 570  & 7&13&10.36 \\
Karate &  Zachary \& Wayne (1977) &  34 & 2 & 78 &  1&17&4.59     \\
Dolphins & Lusseau {\it et al}. (2003)& 62 &2  & 159  &1 &12& 5.12\\
Polbooks  &Krebs (unpublished)  &   92 & 2& 374  &1&24&8.13 \\
UKfaculty & Nepusz {\it  et al}. (2008) &  79 & 3 & 552  &2&39&13.97 \\ 
\end{tabular}
}
\caption{The $8$ network data sets we analyze in this paper. Note that $d_{max}/d_{min}$ can be as large as a few hundreds, suggesting a  severe degree heterogeneity ($d_{min}$, $d_{max}$ and $\bar{d}$ stand for the minimum degree, maximum degree, and average degree, respectively).} 
\end{table}

Conceivably, for some of the data sets, some nodes may have mixed memberships \cite{airoldi2009mixed,MSCORE, JiZhuMM}.  To alleviate the effect,  we did some data pre-processing as follows.   
For the Polbooks data set, we removed all the books that are labeled as ``neutral.''  For  the football data  set, we removed the 5 ``independent" teams. For  the UKfaculty data  set, we removed the smallest group which only contains 2 nodes. 
After the pre-processing, our assumption of  ``no mixed-memberships" is reasonable.  

Natural networks have some noteworthy features. 
\begin{itemize} 
\item {\it Node sparsity and severe degree heterogeneity}.  Take Table \ref{tab:realdata0} for example, even for 
networks with only $1222$ nodes, the degrees for some nodes can be as large as $351$ times higher than those of the others. If we measure the sparsity of a node by its degree, then the sparsity level 
may range significantly from one node to  another. 
\item {\it  Overall network sparsity}. Some networks are much  sparser  than others, and 
the overall network sparsity may range significantly from one network to another. 
\item {\it Weak signal}.  In many cases, the community structures are subtle and masked by strong noise,  where the  signal-to-noise  ratio (SNR) is relatively low. 
\end{itemize}  
It is desirable to have a model that is flexible enough to capture all these features. This is where the DCBM comes in.

%%%%%%%%%%%%
%%%%%%%%%%%%
%%%%%%%%%%%%
\subsection{The degree-corrected block model (DCBM)} \label{subsec:DCBM} 
DCBM is one of the most popular models in network analysis (see for example \cite{DCBM}).   For each node $i$, we encode the community label by a $K$-dimensional vector $\pi_i$, such that for all $1 \leq i \leq n$ and $1 \leq k \leq K$,  
\begin{equation} \label{model1a} 
\mbox{$i \in {\cal C}_k$ if and only if all entries of $\pi_i$ are $0$ except  that  the $k$th entry is $1$}.   
\end{equation} 
In DCBM,  for a matrix $P \in \mathbb{R}^{K,K}$ and parameters $\theta_1, \theta_2, \ldots, \theta_n$,  we  assume the upper triangle of the adjacency matrix $A$  contains  independent Bernoulli random variables satisfying 
\begin{equation} \label{model1b} 
\mathbb{P}(A(i,j) = 1) = \underbrace{\theta_i  \cdot \theta_j}_{\mbox{node specific}} \times  \underbrace{\pi_i' P \pi_j}_{\mbox{community specific}}.  
\end{equation}  
Here, $P$ is a symmetrical and (entry-wise) non-negative matrix that models the community structure and $\theta_1, \theta_2, \ldots, \theta_n$ are positive parameters that model the degree heterogeneity. For identifiability, we assume 
\begin{equation} \label{model1d} 
\mbox{all diagonal entries of $P$ equal to  $1$}.  
\end{equation}

Writing $\theta = (\theta_1, \theta_2, \ldots, \theta_n)'$,  $\Theta = \diag(\theta_1, \theta_2, \ldots, \theta_n)$,  and $\Pi = [\pi_1, \pi_2, \ldots, \pi_n]'$, 
we define
\begin{equation} \label{model1c} 
\Omega = \Theta \Pi P \Pi'\Theta. 
\end{equation} 
Note that when $i \neq j$,  $\Omega(i,j)$ denotes the probability $\mathbb{P}(A(i,j)=1)$. 
Let $W \in \mathbb{R}^{n,n}$ be the {\it centered Bernoulli noise} matrix such that 
$W(i,j) = A(i,j) - \Omega(i,j)$ when $i \neq j$ and $W(i,j) = 0$ if $i = j$.  
We have 
\begin{equation}  \label{lowrankMod}
A = \Omega - \diag(\Omega)   + W  = ``\mbox{main signal}" +  ``\mbox{secondary signal}" + ``\mbox{noise}",   \footnote{We model $\mathbb{E}[A]$ by $\Omega - \diag(\Omega)$ instead of $\Omega$ because the diagonals of $\mathbb{E}[A]$ are all $0$. Here, ``main signal", ``secondary signal", and ``noise" refers to $\Omega$, $-\diag(\Omega)$ and $W$ respectively.} 
\end{equation} 
where $\diag(\Omega)$ stands for the diagonal matrix $\diag(\Omega(1,1), \Omega(2,2), \ldots, 
\Omega(n,n))$. Note that the rank of $\Omega$ is $K$, so 
\eqref{lowrankMod} is a low-rank matrix model.

 In the special case of 
\begin{equation} \label{SBM}  
\theta_1 = \theta_2 = \ldots  = \theta_n = \alpha, \footnote{For SBM, the diagonal entries of $P$ can be unequal. DCBM has more free parameters, so we have to assume that $P$ has unit diagonal entries to maintain identifiability.} 
\end{equation} 
DCBM reduces to the stochastic block  model (SBM).  Note that SBM does not model severe degree heterogeneity. The DCBM is also similar to that in \cite{Chaudhuri} in some sense.

In DCBM, we allow $(\theta, \Pi, P)$ to depend on $n$ so the model is flexible enough to capture all the three features aforementioned of natural networks. 
\begin{itemize} 
\item A reasonable metric for degree heterogeneity is $\theta_{max}/\theta_{min}$, which is allowed to be  large in DCBM.  See Table \ref{tab:realdata0}. 
\item A reasonable metric for overall network sparsity is $\|\theta\|$, and in DCBM, 
$\|\theta\|$ depends on $n$ and is allowed to range freely between $1$ and $\sqrt{n}$ (up to some multi-$\log(n)$ terms),\footnote{A multi-$\log(n)$ term is a term $L_n > 0$ that satisfies $L_n  n^{-\delta} \goto 0$ for any fixed constant $\delta > 0$}  
corresponding to the most sparse networks and the most dense networks, respectively. 
\item A reasonable metric for SNR is $\lambda_K/ \sqrt{\lambda_1}$ (see \cite{JKL2021} for an explanation), where 
$\lambda_k$ is the $k$th largest (in magnitude) eigenvalue of $\Omega$. 
If we allow $\theta$, $P$, and $\Pi$ to depend on $n$, then DCBM is adequate for modeling the weak signal cases where $|\lambda_k|$ may be much smaller than $|\lambda_1|$,  $1 < k \leq K$.  
\end{itemize} 
In many recent works on community detection, it was assumed that the first $K$ eigenvalues are at the same magnitude.  For example, some of these works considered a DCBM model where in (\ref{model1c}), we take 
$P = \alpha_n P_0$.   
Here, $\alpha_n$ is a scaling parameter that may vary with $n$ and $P_0$ being a fixed matrix that  does not vary with $n$.   In this special case,  by similar calculations as in \cite{SCORE}, it is seen that all eigenvalues of $\Omega$ are at the same order under mild regularity conditions on $(\Theta, \Pi)$ (e.g., the $K$ communities are balanced; see \cite{SCORE} for details).      
 Such models do not allow for weak signals, and so are relatively restrictive.

Motivated by these  observations,  it is desirable to have community detection algorithms that  
\begin{itemize} 
\item accommodate severe degree heterogeneity, 
\item are adaptive to different levels of overall network sparsity, 
\item are effective not only for strong signals but also for  weak signals. 
\end{itemize} 

%%%%%%%%%%%%%%%%%
%%%%%%%%%%%%%%%%%
%%%%%%%%%%%%%%%%%
\subsection{The orthodox SCORE}  \label{subsec:SCORE}
SCORE, or {\bf S}pectral {\bf C}lustering {\bf O}n {\bf R}atios-of-{\bf E}igenvectors, is a recent approach to community detection proposed by Jin \cite{SCORE}. SCORE consists of three steps.

{\bf Orthodox SCORE}.  Input: adjacency matrix $A$ and the number of communities $K$.  Output: community labels of all nodes.  
%%%%%%%%
%%%%%%%%
%%%%%%%%
\begin{itemize}
\item 
{\it (PCA).}  Obtain the first $K$ leading eigenvectors $\hat{\xi}_1,\hat{\xi}_2, \ldots,\hat{\xi}_K$ of $A$ (we call $\hat{\xi}_k$  the $k$th leading eigenvector   if the corresponding eigenvalue is the   $k$th largest in  absolute value). 
\item 
{\it (Post-PCA normalization).}  Obtain the $n \times (K-1)$ {\it matrix of entry-wise eigen-ratios} by
\begin{equation} \label{RofSCORE}
\biggl[\frac{\hat{\xi}_2}{\hat{\xi}_1}, \frac{\hat{\xi}_3}{\hat{\xi}_1}, \ldots,  \frac{\hat{\xi}_K}{\hat{\xi}_1}\biggr],   
\end{equation} 
where the ratio of two vectors should be understood as {\it the vector of entry-wise ratios}. \footnote{For example, $\frac{\hat{\xi}_2}{\hat{\xi}_1}$ is the $n$-dimensional vector 
$(\frac{\hat{\xi}_2(1)}{\hat{\xi}_1(1)},  
\frac{\hat{\xi}_2(2)}{\hat{\xi}_1(2)}, \ldots, \frac{\hat{\xi}_2(n)}{\hat{\xi}_1(n)})'$. Note that we may  choose to  threshold all entries of the $n \times (K-1)$ matrix by $\pm \log(n)$ from top and bottom \cite{SCORE}, but this is not always necessary. For all data sets in this paper, 
thresholding or not only has a negligible difference.} 
\item {\it (Clustering)}. Cluster by applying $k$-means to rows of $\hat{R}$, assuming there are $\leq K$ clusters. 
\end{itemize}

Compared to classical spectral clustering, the main innovation of SCORE is the post-PCA normalization. The goal of this step is to mitigate the effect of degree heterogeneity. The degrees contain very little information of the community structure and pose merely as a nuisance, but severe degree heterogeneity makes  
different entries of the leading eigenvectors  badly scaled. As a result, without this step, SCORE tends to cluster nodes according to their degrees instead of the community structure, and thus have unsatisfactory clustering results.  Take the Weblog data for example: with and without this step,  the error rates of SCORE are $58/1222$ and $437/1222$ respectively. 
See \cite{SCORE}  for more discussions.

SCORE is conceptually simple, easy to use, and does not need tuning. In \cite{SCORE, SCC-JiJin}, 
SCORE was shown to be competitive in clustering accuracy.   For computational time, note that in the k-means clustering step of SCORE,  one usually uses the Llyod's algorithm \cite{HTF}; and as a result, SCORE is computationally fast and 
is able to work efficiently for large networks.  See \cite{SCORE} and also Table \ref{tab:experiment}  of the current paper for more discussions.

SCORE is a flexible idea, and can be conveniently extended to many different settings such as network mixed membership estimation \cite{MSCORE}, topic estimation in text mining \cite{Topic}, state aggregation in control theory and reinforcement learning \cite{duan2019state}, analysis of hyper graphs  \cite{ke2019community}, and matrix factorization in image processing.

\subsection{Contribution of this paper}  
For the three features aforementioned, SCORE accommodates   
severe degree heterogeneity and is adaptive to different levels of 
overall network sparsity. However, when it comes to weak signals, 
there are at least two problems that are not answered. 
\begin{itemize} 
\item What is the theoretical behavior of SCORE in the presence of weak signals?  
\item In challenging application problems where the SNR  is  small, is it possible to 
improve SCORE to have better real data performance, without sacrificing its good properties above?  
\end{itemize} 
Note that in the literature, the theoretical analysis on SCORE has been largely focused on the case where the signals are relatively strong; see for example \cite{SCORE}.   

In this paper, we analyze SCORE theoretically, especially for  the weak signal settings. 
We show that for a broad class of settings where we allow  for  weak signals, severe degree heterogeneity,   
and a wide range of overall network sparsity, SCORE  attains an exponential rate of convergence for the Hamming error. We also show that, when the SNR is appropriately large, SCORE  fully recovers the community 
labels except for a small probability. The proof uses  the most recent advancement on entry-wise bounds (a kind of large-deviation bounds) for the  
leading eigenvectors of the adjacency matrix \cite{abbe2017entrywise, MSCORE}.

The theoretical analysis here assures that 
SCORE continues to work well for weak signal settings. This of course 
does not rule out the possibility that a further improved version 
may perform better in real data analysis.  

As a second contribution of the paper, we propose SCORE+ as an improved version of SCORE, 
especially for networks with weak signals. 
We compare SCORE+ with SCORE and several other recent algorithms  using the $8$ data sets in Table \ref{tab:realdata0}. 
For the $6$ data sets where the signals are relatively strong (the clustering 
errors of all methods considered are relatively low), SCORE+ and SCORE have 
comparable performance. For  the $2$ data sets (Simmons and Caltech) where the 
signals are relatively weak (the clustering errors of all methods considered 
  are relatively high),  SCORE+ improves SCORE significantly, and  has  the lowest 
error rates among all methods considered in the paper.  

SCORE+ proposes several changes to SCORE. We carefully explain the 
rationale underlying each of these changes, using a mixture of 
theoretical and numerical study. A much deeper understanding requires 
advanced  
tools in  random  matrix  theory that have not yet been developed,  so 
we leave the study along this line to the future.

\subsection{Content and notations}  
In Section \ref{sec:theory}, we analyze the orthodox SCORE with some new theoretical results.  We show that SCORE attains exponential rates in Hamming clustering errors and  achieves perfect clustering  provided that the SNR is reasonably large. 
In Section \ref{sec:SCOREplus}, we propose SCORE+ as an improved version of SCORE. 
We compare the performance of SCORE+ with SCORE and several   recent approaches 
on community detection using the $8$ data sets in Table \ref{tab:realdata0},  and show that 
SCORE+ has the best  overall  performance.   SCORE+ proposes several changes to SCORE. 
We explain the rationale underlying each of these changes, and especially 
why SCORE+ is expected to have better performance than SCORE for networks with weak signals.   
Section \ref{sec:proof} proves the main results in Section \ref{sec:theory}.  
  
In this paper, for any numbers $\theta_1, \theta_2, \ldots, \theta_n$,  $\theta_{max} = \max\{\theta_1, \theta_2, \ldots, \theta_n\}$,  and $\theta_{min} = \min\{\theta_1, \theta_2, \ldots, \theta_n\}$.  Also, 
$\diag(\theta_1, \theta_2, \ldots, \theta_n)$ denotes the $n \times n$ diagonal matrix with $\theta_i$ being the $i$-th diagonal entry, $1 \leq i \leq n$,     For any vector $a \in \mathbb{R}^n$, $\|a\|_q$ denotes the Euclidean $\ell^q$-norm, and 
we write $\|a\|$ for simplicity when $q = 2$. For any matrix $P \in \mathbb{R}^{n,n}$, 
$\|P\|$ denotes the matrix spectral norm, and $\|P\|_{\mbox{max}}$ 
denotes the maximum $\ell^2$-norm of all the rows of $P$.  For two positive sequences $\{a_n\}$ and $\{b_n\}$, we say $a_n \asymp b_n$ if there are constants $c_2 > c_1 > 0$ such that 
$c_1 a_n \leq b_n \leq c_2 a_n$ for sufficiently large $n$.

%%%%%%%%%%%%%%%
%%%%%%%%%%%%%%%
%%%%%%%%%%%%%%%
\section{SCORE: exponential rate and perfect clustering} \label{sec:theory} 
We provide new theoretical results for the orthodox SCORE, which significantly improves those in \cite{SCORE}. For the ``weak signal" case, the theory in \cite{SCORE} is not applicable but our theory applies. For the ``strong signal" case, compared with \cite{SCORE}, our theory provides a faster rate of convergence for the clustering error and weaker conditions for perfect clustering. 

Consider a sequence of DCBM indexed by $n$, where $(K, \theta,\Pi, P)$ are all allowed to depend on $n$.  Suppose, for a constant $c_1>0$, 
%%%%%%%%%%%%
%%%%%%%%%%%%
%%%%%%%%%%%%
\begin{equation} \label{condition1a} 
\|P\|_{\max}\leq c_1, \qquad \|\theta\| \goto \infty, \qquad \mbox{and} \qquad \theta_{\max} \leq c_1.  
\end{equation}  
Recall that ${\cal C}_1, \ldots, {\cal C}_K$ denote the $K$ true communities. 
For $1 \leq k \leq K$,  let  $n_k = |{\cal C}_k|$ be the size of community $k$,  and let $\theta^{(k)} \in \mathbb{R}^n$ be the vector 
such that $\theta^{(k)}_i = \theta_i$ if $i \in {\cal C}_k$ and 
$\theta^{(k)}_i = 0$ otherwise. We assume, for a constant $c_2>0$,   
%%%%%%%%%%%%%
%%%%%%%%%%%%%
%%%%%%%%%%%%%
\begin{equation} \label{condition1b} 
\max_{1 \leq k \leq K} \{n_k\} \leq c_2 \min_{1 \leq k \leq K} \{n_k\},   
\qquad \mbox{and} \qquad 
\max_{1 \leq k \leq K} \{\|\theta^{(k)}\|\} \leq c_2 \min_{1 \leq k \leq K} \{\|\theta^{(k)}\|\}. 
\end{equation} 
Introduce a diagonal matrix $G \in \mathbb{R}^{K,K}$ by 
\[ 
G=K\|\theta\|^{-2}\cdot \mathrm{diag}(\|\theta^{(1)}\|^2, \|\theta^{(2)}\|^2,\ldots,\|\theta^{(K)}\|^2).
\] 
Let $\mu_k$ denote the $k$th largest eigenvalue (in magnitude) of the $K\times K$ matrix $G^{1/2}PG^{1/2}$, and let $\eta_k$ denote the corresponding eigenvector. We assume, for a constant $c_3\in (0,1)$ and $c_4>0$,  
\beq  \label{condition1c} 
\min_{2\leq k\leq K}|\mu_1-\mu_k|\geq c_3|\mu_1|, \quad\mbox{and } \mbox{$\eta_1$ is a positive vector such that } \frac{\max_{1\leq k\leq K}\{\eta_1(k)\}}{\min_{1\leq k\leq K}\{\eta_1(k)\}}\leq c_4. 
\eeq

These conditions are mild. For \eqref{condition1a}, recall that $\|\theta\|$ measures the overall network sparsity, and the interesting range of $\|\theta\|$ is between $1$ and $\sqrt{n}$, up to some multi-$\log(n)$ terms (see footnote 3). Therefore, it is mild to assume $\|\theta\|\to\infty$. Condition \eqref{condition1b} requires that the communities are balanced in size and in total squared degrees, which is mild.

Condition \eqref{condition1c} is also mild. The most challenging case for network analysis is when the matrix $P$ gets very close to the matrix of all ones, where it is hard to distinguish one community from 
another.  The condition  only rules out the less relevant cases such as the network is disconnected or approximately so. These are strong 
signal cases where it is relatively easy to distinguish one community from another.  
Note that the condition is satisfied if all entries of $P$ are lower bounded by a constant or if $K$ is fixed and $P$ converges to a fixed irreducible matrix (see Section A.2 of \cite{MSCORE} for a discussion of this condition). 

In a hypothesis testing framework, \cite{JKL2021} has pointed out that a reasonable metric for SNR in a DCBM is $|\lambda_K| / \sqrt{\lambda_1}$, where we recall that $\lambda_k$ denotes the $k$th largest eigenvalue (in magnitude) of $\Omega=\Theta\Pi P\Pi'\Theta$ and that $\lambda_1$ is always positive \cite{SCORE, MSCORE}. 
We also introduce a quantity to measure the severity of degree heterogeneity:
\[
\alpha(\theta) = (\theta_{\min} / \theta_{\max}) \cdot (\|\theta\|/\sqrt{\theta_{\max} \|\theta\|_1}) \quad\in\quad (0,1]. 
\] 
The smaller $\alpha(\theta)$, the more severe the degree heterogeneity. When $\theta_{\max}\asymp \theta_{\min}$, $\alpha(\theta)$ is bounded below from $0$ by a constant. In the presence of severe degree heterogeneity, $\alpha(\theta)$ gets close to $0$. 
We shall see that the clustering power of SCORE depends on
\beq \label{SNR}
s_n = \alpha(\theta)\cdot (|\lambda_K| / \sqrt{\lambda_1}), 
\eeq
which is a combination of the SNR and the severity of degree heterogeneity.

Let $\widehat{\Pi}=[\hat{\pi}_1,\hat{\pi}_2,\ldots,\hat{\pi}_n]'$ denote the matrix of estimated community labels by the orthodox SCORE. 
Define the Hamming error rate (per node)  for clustering  by 
\[
\mathrm{Hamm}(\widehat{\Pi},\Pi) = \frac{1}{n} \sum_{i=1}^n 1\{\hat{\pi}_i\neq \pi_i\}, \qquad \mbox{up to a permutation on columns of $\widehat{\Pi}$}. 
\]
The next theorem is proved in Section~\ref{sec:proof}. 
%%%%%%%%%%%
%%%%%%%%%%%
%%%%%%%%%%
\begin{thm}\label{thm:Hamming}
Consider a sequence of DCBM indexed by $n$, where \eqref{condition1a}-\eqref{condition1c} hold. Let $s_n$ be as in \eqref{SNR}. There exist appropriately small constants $a_1, a_2>0$, which depend on the constants $c_1$-$c_4$ in the regularity conditions, such that, as long as $s_n\geq a^{-1}_1K^4\sqrt{\log(n)}$, for sufficiently large $n$:
\[
\mathbb{E}\big[\mathrm{Hamm}(\widehat{\Pi},\Pi) \big] \leq  \frac{2K}{n}\sum_{i=1}^n \exp\left( - a_2 \theta_i\cdot \min\biggl\{ \frac{(|\lambda_K|/\sqrt{\lambda_1})^2\|\theta\|^2}{K^2\|\theta\|_3^3},\; \frac{(|\lambda_K|/\sqrt{\lambda_1})\|\theta\|}{K\theta_{\max}}  \biggr\}\right)+o(n^{-3}).  
\]
\end{thm}

Theorem~\ref{thm:Hamming} implies that the Hamming clustering error of SCORE depends on $(\theta_1,\ldots,\theta_n)$ in an exponential form. This significantly improves the bound in \cite{SCORE}, which depends on $(\theta_1,\ldots,\theta_n)$ in a polynomial form. Additionally, Theorem~\ref{thm:Hamming} suggests that the nodes with smaller $\theta_i$ have large contributions to the Hamming clustering error, i.e., it is more likely for the algorithm to make errors on low-degree nodes. 

The clustering error has an easy-to-digest form in special examples. 
%%%%%%%%%
%%%%%%%%%
%%%%%%%%%%%%%%
\begin{cor} \label{cor:Hamm}
Consider a special DCBM, where 
\[
P = (1-b)I_K + b{\bf 1}_K{\bf 1}_K', \qquad \pi_i\overset{iid}{\sim}\mathrm{Uniform}(\{e_1,e_2,\ldots,e_K\}). 
\] 
Suppose $K$ is fixed and $\theta$ satisfies that $\theta_{\max}\leq C\theta_{\min}$. There exist appropriately small constants $\tilde{a}_1,\tilde{a}_2>0$, such that, as long as $(1-b)\|\theta\|\geq a_1^{-1}K^4\sqrt{\log(n)}$, for sufficiently large $n$,
\[
\mathbb{E}\big[\mathrm{Hamm}(\widehat{\Pi},\Pi) \big] \leq \frac{2K}{n} \sum_{i=1}^n \exp\left(-a_2\frac{\theta_i}{\bar{\theta}}\cdot \frac{(1-b)^2\|\theta\|^2}{K^3} \right) + o(n^{-3}). 
\]
\end{cor}

In this special example, $\|\theta\|^2$ characterizes the average node degrees, and $(1-b)$ captures the ``similarity" across communities. The clustering power of SCORE is governed by $
s_n \asymp (1-b)\|\theta\|^2$. 
The bound in Corollary~\ref{cor:Hamm} matches with the minimax bound in \cite{gao2018community}, except for the constant $2K$ in front and the constant $a_2$ in the exponent. \footnote{When translating the bound in \cite{gao2018community}, we notice that $\theta_i$ there have been normalized, so that their $\theta_i$ corresponds to our $(\theta_i/\bar{\theta})$.} It was shown in \cite{gao2018community} that the exponential error rate can be attained by first applying spectral clustering and then conducting a refinement, where the refinement step was motivated by technical convenience. In fact, numerical studies suggest that spectral clustering alone can attain exponential error rates. Theorem~\ref{thm:Hamming} and Corollary~\ref{cor:Hamm} provide a rigorous theoretical justification.

The next theorem states that SCORE can exactly recover the community labels with high probability, provided that the SNR is appropriately large.  
\begin{thm}\label{thm:main}
Consider a sequence of DCBM indexed by $n$, where \eqref{condition1a}-\eqref{condition1c} hold. Let $s_n$ be as in \eqref{SNR}. If $s_n \geq CK^4\sqrt{\log(n)}$ for a sufficiently large constant $C>0$, then we have that, up to a permutation on columns of $\widehat{\Pi}$, 
\[
\mathbb{P}(\widehat{\Pi} \neq \Pi) = o(n^{-3}).  
\]
Furthermore, if $K$ is finite and $\theta_{\max}\leq C\theta_{\min}$, then the above is true as long as $|\lambda_K|/\sqrt{\lambda_1}$ exceeds a sufficiently large constant. 
\end{thm}
\noindent

The condition on $s_n$ cannot be significantly improved. 
Take the case of fixed $K$ and with moderate degree heterogeneity ($\theta_{\max}\leq C\theta_{\min}$) for example. In this case, 
the condition becomes $|\lambda_K| / \sqrt{\lambda_1} \geq C$ for a large enough constant $C>0$.  
It was shown in \cite{JKL2021} that, if  we allow $|\lambda_K| / \sqrt{\lambda_1} \goto 0$, 
then we end up with a class of models that is too broad so we can find 
two sequences of DCBM models with different (fixed) $K$ but are indistinguishable from each other. 
In such settings, successful clustering is impossible. 
 
In the literature, the exponential error rate and the perfect clustering property were mainly obtained for non-spectral methods (e.g., \cite{gao2018community,chen2015convexified}). While spectral methods are practically popular, its theoretical analysis is challenging, since it requires sharp entry-wise bounds for eigenvectors. A few existing works either focus on SBM which does not allow for degree heterogeneity (e.g., \cite{abbe2017entrywise,su2019strong}) or restrict to the ``strong signal" case and dense networks (e.g., \cite{liu2019community}). Our results are new as we provide the first exponential rate result and perfect clustering result for spectral methods that accommodate severe degree heterogeneity, sparse networks, and weak signals.

Our analysis uses some results on the spectral analysis of the adjacency  matrices from  our work \cite{MSCORE}, especially the entry-wise large-deviation bounds for empirical eigenvectors. We refer interesting readers to a detailed description in \cite{MSCORE}.  It is understood that the main technical difficulty of analyzing spectral methods lies in entry-wise analysis of eigenvectors. Recent progress includes (but does not limit to)   
\cite{abbe2017entrywise,fan2019simple,MSCORE, liu2019community,mao2020estimating,su2019strong}.

%%%%%%%%%%%%%%%
%%%%%%%%%%%%%%%
%%%%%%%%%%%%%%%
\section{SCORE+, a refinement especially for weak signals} \label{sec:SCOREplus}
We propose SCORE+  as a refinement of SCORE for community detection.  SCORE+ inherits the appealing features of SCORE. It improves the performance of SCORE in real applications,  especially for networks with weak signals.

\subsection{SCORE+} \label{subsec:SCOREplus} 
Recall that under DCBM,  
\[
A = \Omega - \diag(\Omega) +W = ``\mbox{main signal}" +  ``\mbox{secondary signal}" + ``\mbox{noise}",     
\]  
where  the ``main signal" matrix $\Omega$ equals to $\Theta \Pi P \Pi' \Theta$ and has a rank $K$. SCORE+ is motivated by several observations about SCORE. 
\begin{itemize}  
\item Due to severe degree heterogeneity, different rows of  the  ``signal" matrix  and the ``noise" matrix  are in very different scales. We need two normalizations: a pre-PCA 
normalization to mitigate the effects of degree heterogeneity on the ``noise" matrix, and a post-PCA 
normalization (as in SCORE) on the ``signal" matrix; we find  that  an appropriate pre-PCA normalization is  Laplacian regularization.\footnote{This is analogous to the Students' $t$-test, where for $n$ samples from an unknown distribution, the $t$-test uses a normalization for the mean  and a normalization for the variance.}    
See Section \ref{subsec:explanation1} for more explanations.
\item The idea of PCA is dimension reduction: We project rows of $A$ to the $K$-dimensional space spanned by the first $K$ eigenvectors of $A$,  $\hat{\xi}_1, \hat{\xi}_2, \ldots, \hat{\xi}_K$,  and reduce $A$ to the $n \times K$ matrix of projection coefficients:  
\[
[\hat{\eta}_1, \hat{\eta}_2, \ldots, \hat{\eta}_K] \equiv  [\hat{\xi}_1, \hat{\xi}_2, \ldots, \hat{\xi}_K]  \cdot \diag(\hat{\lambda}_1, \hat{\lambda}_2, \ldots, \hat{\lambda}_K).    
\] 
Therefore, in SCORE, it is better to apply the post-PCA normalization to  $[\hat{\eta}_1, \hat{\eta}_2, \ldots, \hat{\eta}_K]$ instead of $[\hat{\xi}_1, \hat{\xi}_2, \ldots, \hat{\xi}_K]$;  the two post-PCA normalization matrices (old and new) satisfy 
\[
\biggl[\frac{\hat{\eta}_2}{\hat{\eta}_1}, \frac{\hat{\eta}_2}{\hat{\eta}_1}, \ldots, \frac{\hat{\eta}_K}{\hat{\eta}_1} \biggr] =  \biggl[\frac{\hat{\xi}_2}{\hat{\xi}_1}, \frac{\hat{\xi}_2}{\hat{\xi}_1}, \ldots, \frac{\hat{\xi}_K}{\hat{\xi}_1} \biggr]  \cdot  \diag\biggl(\frac{\hat{\lambda}_2}{\hat{\lambda}_1}, \frac{\hat{\lambda}_3}{\hat{\lambda}_1}, \ldots,  
\frac{\hat{\lambda}_K}{\hat{\lambda}_1}\biggr). 
\]  
In effect, the new change is using eigenvalues to re-weight the columns of  $\bigl[\frac{\hat{\xi}_2}{\hat{\xi}_1}, \frac{\hat{\xi}_2}{\hat{\xi}_1}, \ldots, \frac{\hat{\xi}_K}{\hat{\xi}_1}\big]$.  
See Section \ref{subsec:explanation2}  for more explanations. 
\item In SCORE, we only use the first $K$ eigenvectors for clustering, which is reasonable in the ``strong signal" case, where all the nonzero eigenvalues of the ``signal" matrix are much larger than the spectral norm of the ``noise" matrix (in absolute value).  In the ``weak signal" case, some nonzero eigenvalues of the ``signal" can be smaller than the spectral norm of the ``noise",  and we may need one or more additional eigenvectors of $A$ for clustering. In Section \ref{subsec:explanation3}, we have an in-depth study on the weak signal case; see details therein.  
\end{itemize}

{\bf SCORE$+$}.  Input: $A$, $K$, a ridge regularization parameter $\delta > 0$ and a threshold $t > 0$. Output: class labels for all $n$ nodes.  
\begin{itemize}
\item {\it (Pre-PCA normalization with Laplacian).} Let $D = \diag(d_1, d_2, \ldots, d_n)$ where $d_i$ is the degree of node $i$.   Obtain the graph Laplacian with 
ridge regularization by 
\[
L_{\delta} = (D + \delta \cdot d_{max} \cdot  I_n)^{-1/2} A (D + \delta \cdot d_{max} \cdot  I_n)^{-1/2}, \qquad \mbox{where $d_{max} = \max_{1 \leq i \leq n} \{d_i\}$}. 
\]
Note that 
the ratio between the largest diagonal entry of $D + \delta d_{max} I_n$ and the smallest one  is 
smaller than  $(1 + \delta)/\delta$. Conventional choices of $\delta$ are $0.05$ and $0.10$. 
\item {\it (PCA, where we retain possibly an additional eigenvector).}  
We assess the aforementioned ``signal weakness" by $1 - [\hat{\lambda}_{K+1}/\hat{\lambda}_K]$, and include an additional eigenvector for clustering if and only if 
\[
(1 - [\hat{\lambda}_{K+1} /\hat{\lambda}_{K}]) \leq t, \qquad (\mbox{a  conventional choice of $t$ is $0.10$}).  
\]
\item {\it (Post-PCA normalization).}  Let $M$ be the number of eigenvectors we decide in the last step (therefore, either $M = K$ or $M = K +1$).  Obtain the matrix of entry-wise eigen-ratios  by
\begin{equation} \label{newSCORE}
\hat{R} = \biggl[\frac{\hat{\eta}_2}{\hat{\eta}_1},  \frac{\hat{\eta}_3}{\hat{\eta}_1}, \ldots,  \frac{\hat{\eta}_M}{\hat{\eta}_1}\biggr],   \qquad \mbox{where $\hat{\eta}_k = \hat{\lambda}_k \hat{\xi}_k$, $1 \leq k \leq M$}.   
\end{equation} 
\item {\it (Clustering)}. Apply classical $k$-means to  the rows of  $\hat{R}$, assuming $\leq K$ clusters. 
\end{itemize}
The code is available at \url{http://zke.fas.harvard.edu/software.html}.

Compared to SCORE, SCORE+ (a) adds a pre-PCA normalization step, (b) may select one more eigenvectors for later use if necessary,     and (c) uses eigenvalues to re-weight the columns of $\bigl[\frac{\hat{\xi}_2}{\hat{\xi}_1}, \frac{\hat{\xi}_3}
{\hat{\xi}_1}, \ldots, \frac{\hat{\xi}_K}{\hat{\xi}_1} \bigr]$.  In Section~\ref{subsec:why}, we further explain the rationale underlying these refinements.  
 
%%%%%%%%%%%%%
%%%%%%%%%%%%%
%%%%%%%%%%%%%
\subsection{Numerical comparisons}  \label{subsec:Numerical} 
We compare SCORE+ with a few recent methods: Orthodox SCORE,  the convexified modularity maximization (CMM) method by \cite{chen2015convexified}, the latent space model based (LSCD) method by \cite{ma2020universal},  the normalized spectral clustering (OCCAM) method for potentially overlapping communities by \cite{JiZhuMM}, and the regularized spectral clustering (RSC) method by \cite{qin2013regularized}.  For each method, we measure the clustering error rate by 
\[
\mathrm{min}_{\mbox{$\{\tau$: permutation over $\{1, 2, \ldots, K\}\}$}} \frac{1}{n} \sum_{i = 1}^n 1\{ \tau(\ell_i) \neq \ell_i\}, 
\] 
where $\ell_i$ and $\hat{\ell_i}$ are the true and estimated labels of node $i$. 

The error rates are in Table \ref{tab:error}, where for SCORE+, we take $(t, \delta) = (0.1, 0.1)$. For the three relatively large networks (Weblog, Simmons, Caltech),
the error rates of SCORE+ are the best among all methods,  and for the other networks, the error rates are close to the best.   Especially,   SCORE+ provides a commendable improvement for the Simmons and Caltech data sets.  In Section \ref{subsec:explanation3},   we show that  the Simmons and Caltech data sets are ``weak signal" networks, and all other networks are ``strong signal" networks. 
%%%%%%%%%%%%%%%      
%%%%%%%%%%%%%%%
%%%%%%%%%%%%%%%
\begin{table}[htb!]  
\centering
\scalebox{.98}{
\begin{tabular}{l |  lll  ll  ll}
Dataset &  SCORE  & SCORE$+$ & CMM  & LSCD& OCCAM&RSC\\
\hline
Weblogs & 58/1222  & \bf{51/1222}  & 62/1222 &58/1222 &65/1222& 64/1222\\
Simmons  & 268/1137 & $\bf{127/1137}$ & 137/1137 & 134/1137 & 266/1137& 244/1137 \\
Caltech   & 183/590 & \bf{98/590} & {124/590} & {106/590}  & 189/590& 170/590 \\
Football  & 5/110 & 6/110 &  7/110&  21/110&  $\bf{4/110}$&  5/110\\
Karate &  $\bf{0/34}$ & 1/34 &   $\bf{0/34}$&  1/34&$\bf{0/34}$&  $\bf{0/34}$\\
Dolphins  &  $\bf{0/62}$ &  {2/62} & 1/62& 2/62& 1/62& 1/62\\
Polbooks  &$\bf{1/92}$ & 2/92 & $\bf{1/92}$& 3/92 & 3/92 & 3/92\\
UKfaculty  & 2/79 & {2/79} & 7/79& 1/79 & {5/79} & $\bf{0/79}$\\
\hline
\end{tabular}
}
\caption{Error rates on the $8$ datasets listed in Table~\ref{tab:realdata0}. For SCORE+, we set $(t,\delta)=(0.1,0.1)$.} \label{tab:error} 
\end{table}

RSC \cite{qin2013regularized} is an interesting method that applies the idea of SCORE to the graph Laplacian. It can be viewed as adding a pre-PCA normalization step to SCORE (but it does not include other refinements as in SCORE+).  For three of the data sets (Simmons, Caltech, UKfaculty), the modification provides a small improvement,  and for three of the data sets (Weblogs, Dolphins, Polbooks), the modification hurts a little bit. 
The performance of OCCAM is more or less similar to that of SCORE and RSC,
which is not surprising, because OCCAM is also a normalized spectral method.

The error rates of CMM and LSCD are comparable with  that of SCORE+ in most data sets, except 
that CMM and LSCD have unsatisfactory results for UKfaculty and Football, respectively. 
For the three small data sets (Karate, Dolphins, Polbooks), the three methods have similar error rates, with CMM being slightly better. For the three large data sets (Weblogs, Simmons, Caltech), SCORE+ is better than LSCD, and LSCD is better than CMM.   

LSCD is an iterative algorithm which solves a non-convex optimization with rank constraint. 
Since the algorithm only provides a local optimum, the difference between this local optimum and the global optimum may be large, especially for large $K$. This partially explains why LSCD performs unsatisfactorily on Football, for which data  set  $K=11$. CMM first solves a convexified modularity maximization problem to get an $n\times n$ matrix $\hat{Y}$ and then applies $k$-median to rows of $\hat{Y}$. The matrix $\hat{Y}$ targets on approximating a rank-$K$ matrix, but for UKfaculty, the output $\hat{Y}$ has a large $(K+1)$th eigenvalue. This partially explains why CMM performs unsatisfactorily on this data set.

SCORE+ has two tuning parameters $(t, \delta)$, but each of which is easy to set, guided by  common sense. Moreover,  SCORE+ is relatively insensitive to the ridge regularization parameter $\delta$: in Table \ref{tab:tuning6}, we investigate SCORE+ by setting $t = 0.10$ and letting $\delta$ range from $0.025$ to $0.2$ with an increment of $0.025$. The results suggest SCORE+ is relatively insensitive to different choices of $\delta$. 
In Section \ref{subsec:explanation3}, we discuss further how to set the tuning parameter $t$.

\begin{table}[htb!]  
\centering
\scalebox{.9}{
\begin{tabular}{l |  c  | c  |  c  | c|c | c  | c| l  }
\hline
$\delta$ &  Polblogs & Karate & Dolphins &Football  & Polbooks &UKfaculty   & Simmons&  Caltech\\
\hline
0.025 & 57 & 1& 0&6&2 &1&127 &99  \\
0.05 & 54 & 1 & 1&6&2&2&$\bf{117}$  & 100   \\
0.075 & $\bf{51}$  &1 & 1&6&2&2& 121 & 99  \\
0.10 &$\bf{51}$  &1 & 2&6&2&2 &127 & $\bf{98}$  \\
0.125 & 53 &1 &3&6&2&2&134 & 101   \\
0.15  & 54 &1 &3&6&2&2&137 &101    \\
0.175  & 56 &0 &3&6&2&3&141 & 104   \\
0.20  & 58 &0 &3&6&2&3&142 &105 \\
\hline
\end{tabular}
}
\caption{Community detection errors of SCORE+ for different $\delta$ ($t$ is fixed at  $0.10$).} \label{tab:tuning6} 
\end{table}

Computationally, SCORE and OCCAM are the fastest,  SCORE+ and RSC are slightly slower (the extra computing time is mostly due to the pre-PCA step),  and CMM and LSCD are significantly slower, especially for large networks. For comparison of computing time, it makes more sense to use networks larger than those in Table \ref{tab:realdata0}. We simulate networks from the DCBM model in Section~\ref{subsec:DCBM}. In a DCBM  with  $n$ nodes and $K$ communities, the upper triangle of $A$ contains independent Bernoulli variables, with 
\[
\mathbb{E}[A] = \Omega  - \diag(\Omega),    \qquad \mbox{and} \qquad \Omega = \Theta \Pi P \Pi' \Theta,   
\] 
where $P$ is a $K \times K$ symmetric nonnegative matrix,   $\Theta = \diag(\theta_1, \theta_2, \ldots, \theta_n)$   
with $\theta_i > 0$ being the degree parameters, and  $\Pi$ is the $n\times K$ label matrix. 
For simulations, we let $n$ range in  $\{1000, 2000, 4000, 7000, 10000\}$, and for each fixed $n$, 
 \begin{itemize} 
\item for $c_n = 3 \log(n)/n$ and $(\alpha, \beta) = (5,4/5)$,   generate $\theta_i$ such that $(\theta_i/c_n)$  are $iid$ from $\mathrm{Pareto}(\alpha,\beta)$;
\item fix $K = 4$ and let $\Pi$ be the matrix where the first, second, third, and last quarter of rows equal to $e_1, e_2, e_3, e_4$, respectively;   
\item consider two experiments, where respectively,  the $P$ matrix  is 
\[
\left[ 
\begin{array}{cccc}
1 & 1/3 & 1/3 & 1/3 \\
1/3 & 1 & 1/3 & 1/3 \\
1/3 & 1/3 & 1 & 1/3 \\
1/3 & 1/3 & 1/3 & 1 \\ 
\end{array} 
\right] 
\qquad 
\mbox{and} 
\qquad 
\left[
\begin{array}{cccc}
1     & 2/3  & .1  & .1  \\
2/3  & 1     & .5  & .5 \\
.1  &  .5 & 1     & .5 \\
.1  &  .5 & .5  & 1\\
\end{array}
\right];
\] 
the value of $|\lambda_K(P)|/\lambda_1(P)$ is $0.333$ for the left and $0.083$ for the right, so that they represent the ``strong signal" case and ``weak signal" case, respectively. 
\end{itemize}  
The error rates and computing time are reported in Table \ref{tab:experiment} (both error rates and computing time are the average of $10$ independent repetitions). 
\begin{table}[htb!]  
\centering
\scalebox{.92}{
\begin{tabular}{crcccccc}
\hline
& $n$ &  SCORE& SCORE$+$ & CMM  & LSCD& OCCAM&RSC\\
\hline
 & 1000 &0.08 (0s) &0.04 (3s)  & $ \mathbf{0.03}$ (32s)  &   0.05 (35s) & 0.06 (0s)& $0.04$ (0s) \\
& 2000 &0.06 (1s) & $\mathbf{0.03}$ (4s) & $\mathbf{0.03}$ (240s) &   0.04 (180s) & 0.07 (1s) &  
$\mathbf{0.03}$ (1s)\\
& 4000 & 0.16 (2s) & $\mathbf{0.02}$ (7s) &  0.03 (1930s) & 0.03 (967s)& 0.14 (2s) & 
$\mathbf{0.02}$ (2s)\\
& 7000 & 0.04 (6s) & $\mathbf{0.01}$ (22s)& 0.02 (10500s) & 0.02(2900s) &  0.05 (6s) & 
0.02 (13s) \\
& 10000 & 0.03 (9s) & $\mathbf{0.01}$ (39s) &$\mathbf{0.01}$ (31000s) & 0.02 (6000s)& 0.03 (9s) & $\mathbf{0.01}$  (21s) \\
\hline
\hline
  & 1000 & 0.37 (1s) & $\mathbf{0.07}$ (3s) &  0.10 (47s) & 0.17 (40s)  & 0.37 (0s)& 0.32 (0s) \\
& 2000 & 0.31 (1s) &$\mathbf{0.05}$ (4s) & 0.07 (313s) &   0.06 (194s) & 0.31 (1s) & 0.32 (1s)\\
& 4000 & 0.30 (1s)& $\mathbf{0.05}$ (7s)&  0.06 (2130s) & 0.06 (960s) &0.30 (1s)&0.30 (2s)\\
& 7000 & 0.26 (4s) &$\mathbf{0.03}$ (22s)&  0.05 (10800s)& 0.05 (2900s) & 0.27 (4s)& 0.28 (13s) \\
&10000 & 0.27 (9s) &$\mathbf{0.03}$ (39s) & 0.04 (32150s)& 0.04 (6100s) & 0.28 (9s)&0.29 (21s)\\
\hline
\end{tabular}
}
\caption{Comparison of error rates and computation time on simulated data. Top: Experiment 1 (``strong signal"). Bottom: Experiment 2 (``weak signal")}. \label{tab:experiment} 
\end{table}

In summary, SCORE+ compares favorably over other methods both in error rates and in computing times, either for networks with ``strong signals" or ``weak signals".

%%%%%%%%%%
%%%%%%%%%%
%%%%%%%%%% 
\subsection{Rationale underlying the key components of SCORE+} \label{subsec:why} 
SCORE+ contains  4 components: the post-PCA normalization that was originally 
proposed in SCORE, and 3 proposed refinements (pre-PCA normalization using the 
Laplacian regularization, reweighing 
the leading eigenvectors by eigenvalues, and recruiting one more eigenvector 
for use  when the eigengap is small).  We now explain the rationale  of  each of these components. 

Recall that under DCBM, 
\[
A = \Omega - \diag(\Omega) +W = ``\mbox{main signal}" +  ``\mbox{secondary signal}" + ``\mbox{noise}",     
\]  
where the ``main signal" matrix $\Omega$ equals to $\Theta \Pi P \Pi' \Theta$ and has a rank $K$. 
Let $\xi_1,\xi_2,\ldots,\xi_K$ be the eigenvectors of $\Omega$ associated with $K$ largest eigenvalues in magnitude. Write $\Xi=[\xi_1,\xi_2,\ldots,\xi_K]=[\Xi_1,\Xi_2,\ldots,\Xi_n]'$. 

%%%%%%%%%%%%
%%%%%%%%%%%%
\subsubsection{Rationale underlying the post-PCA normalization} 
The rationale underlying the post-PCA normalization was carefully explained in \cite{SCORE}, 
so we keep the discussion brief here. 
Under DCBM, Jin \cite{SCORE} observed that 
\[
\Xi_i = \theta_i \cdot q_i, \qquad \mbox{where $\{q_1,\ldots,q_n\}$ take only $K$ distinct values in $\mathbb{R}^K$}. 
\]
Without $\theta_i$'s, we can directly apply $k$-means to rows of $\Xi_i$. Now, with the degree parameters, \cite{SCORE} considered the family of {\it scaling invariant mappings (SIM)}, $\mathbb{M}:\mathbb{R}^K\to\mathbb{R}^K$, such that $\mathbb{M}(ax)=\mathbb{M}(x)$ for any $a>0$ and $x\in\mathbb{R}^K$, and proposed the post-PCA normalization
\[
\Xi_i\qquad \mapsto \qquad \mathbb{M}(\Xi_i), \qquad 1\leq i\leq n. 
\] 
The scaling-invariance property of $\mathbb{M}$ ensures $\{\mathbb{M}(\Xi_1),\ldots,\mathbb{M}(\Xi_n)\}$ take only $K$ distinct values, so that we can apply $k$-means. Two examples of SIM include:
\begin{itemize} \itemsep -2pt
\item $\mathbb{M}(x)=(x_2/x_1,x_3/x_1,\ldots,x_n/x_1)'$, i.e., normalizing $\Xi_i$ by its first entry; 
\item $\mathbb{M}(x)=\|x\|_p^{-1}x$, i.e., normalizing $\Xi_i$ by its $L_p$-norm. 
\end{itemize}
The first one was recommended by \cite{SCORE} and is commonly referred to as SCORE. The second one is a variant of SCORE and was proposed in the supplement of \cite{SCORE}. 

In the more general DCMM model with mixed membership, \cite{MSCORE} discovered that the post-SCORE matrix is associated with a low-dimensional simplex geometry and developed SCORE into a simplex-vertex-hunting method for mixed-membership estimation. Interestingly, although each normalization in the scaling invariant family proposed by  \cite{SCORE} works for DCBM, only the SCORE normalization produces the desired simplex geometry under DCMM.  
 %%%%%%%%%%%%
%%%%%%%%%%%%
%%%%%%%%%%%%
\subsubsection{Why the Laplacian is the right pre-PCA normalization} 
\label{subsec:explanation1} 
%%%%%%%%%%%%%%%%%%%%%%%
The target of SCORE is to remove the effect of degree heterogeneity in the ``main signal" matrix $\Omega$. However, the ``noise" matrix $W=A-\mathbb{E}[A]$ is also affected by degree heterogeneity and requires a proper normalization. We note that, since PCA only retains a few leading eigenvectors which are driven by ``signal," the ``noise" is largely removed after conducing PCA. Therefore, one has to use a pre-PCA operation to normalize the ``noise" matrix. 
 
Our idea is to re-weight the rows and columns of $A$ by node degrees. Let $D$ be the diagonal matrix where $D(i,i)$ is the degree of node $i$. 
There are many ways for pre-PCA normalization, and simple choices include 
\begin{itemize} \itemsep -1pt
\item $A \mapsto D^{-1/2} A D^{-1/2}$. 
\item $A \mapsto D^{-1} A D^{-1}$.  
\end{itemize} 
Which one is the right choice? 

Given an arbitrary positive diagonal matrix $H$, write 
\[
H^{-1} A H^{-1} = \underbrace{H^{-1} \Omega H^{-1}}_{``signal"} + \underbrace{H^{-1}[A-\mathbb{E}[A]-\diag(\Omega)] H^{-1}}_{``noise"}.  
\] 
The best pre-PCA normalization is such that, despite severe degree heterogeneity, the variances of all entries of the ``noise" matrix are at the same order \cite{jin2018optimal}. Under DCBM, by direct calculations, 
\[
\mbox{variance of $(i,j)$-entry of ``noise"} \asymp \frac{\theta_i\theta_j}{h_i^2h_j^2} \qquad\quad\Longrightarrow \qquad\quad \mbox{we hope}\quad h_i\propto \sqrt{\theta_i}. 
\] 
At the same time, the node degrees satisfy
\[
d_i\qquad\propto \qquad \theta_i,  \qquad \mbox{approximately}. 
\]
Therefore, the right choice is $h_i\propto \sqrt{d_i}$, i.e., we should use the pre-PCA normalization of $A\mapsto D^{-1/2}AD^{-1/2}$.  See \cite{milena} for a similar finding. For better practical performance, we add a ridge regularization.

Besides normalizing the ``noise" matrix, the pre-PCA normalization also changes the ``signal" matrix from $\Omega$ to $D^{-1/2}\Omega D^{-1/2}$.  Fortunately, the new ``signal" matrix has a similar form as $\Omega=\Theta\Pi P\Pi'\Theta$,  except that  $\Theta$ is  replaced by $D^{-1/2}\Theta$, so the post-PCA normalization of SCORE is still valid.

%%%%%%%%%%%%%%
%%%%%%%%%%%%%%
%%%%%%%%%%%%%%
\subsubsection{Why $\hat{\eta}_k$ is the appropriate choice in post-PCA normalization}
\label{subsec:explanation2}  
In the post-PCA normalization, SCORE+ constructs the matrix of entry-wise eigen-ratios using $\hat{\eta}_1,\ldots,\hat{\eta}_K$, where each $\hat{\eta}_k$ is $\hat{\xi}_k$ weighted by the corresponding eigenvalue. There are many ways of weighting the eigenvectors, and simple choices include
\begin{itemize} 
\item $[\hat{\xi}_1, \hat{\xi}_2, \ldots, \hat{\xi}_K]  \cdot \diag(\hat{\lambda}_1, \hat{\lambda}_2, \ldots, \hat{\lambda}_K)$.  
\item $[\hat{\xi}_1, \hat{\xi}_2, \ldots, \hat{\xi}_K]  \cdot \diag\bigl(\sqrt{\hat{\lambda}_1}, \sqrt{\hat{\lambda}_2}, \ldots, \sqrt{\hat{\lambda}_K}\bigr)$.  
\end{itemize} 
Why do we choose the first one?

We briefly explained  it in Section~\ref{subsec:SCOREplus} using the perspective of projecting rows of data matrix to the span of $\hat{\xi}_1,\ldots,\hat{\xi}_K$. We now take a different perspective. Recall that $L_\delta$ is the  regularized graph Laplacian, by \cite{abbe2017entrywise}, the first-order approximations of eigenvectors are 
\[
\hat{\xi}_k \approx  \frac{1}{\lambda_k} L_\delta\xi_k\approx \xi_k +  \frac{1}{\lambda_k} (L_\delta-\mathbb{E}[L_\delta]) \xi_k. 
\]  
Intuitively speaking, since each $\xi_k$ has a unit-norm, the ``noise" vector $(L_\delta-\mathbb{E}[L_\delta]) \xi_k$ is at the same scale for different $k$; it implies that the noise level in different eigenvectors is proportional to $1/\lambda_k$. This means $\hat{\xi}_1$ is less noisy than $\hat{\xi}_2$, and $\hat{\xi}_2$ is less noisy than $\hat{\xi}_3$, and so on. By weighing the eigenvectors by $\hat{\lambda}_k$, the noise level in $\hat{\eta}_1,\ldots,\hat{\eta}_K$ is approximately at the same order.  

In most theoretical studies, $\lambda_1,\ldots,\lambda_K$ are assumed at the same order, so whether or not to re-weight the eigenvectors does not affect the rate of convergence. However, in many real data, the magnitudes of the first a few eigenvalues can be considerably different, so such a weighting scheme does improve the numerical performance.

%%%%%%%%%%%%%
%%%%%%%%%%%%%
%%%%%%%%%%%%%
\subsubsection{When we should choose one more eigenvector for inference}
\label{subsec:explanation3}  

In SCORE+, we retain $M$ eigenvectors in the PCA step for later uses, where  
\[
M  = 
\begin{cases}
K,  &\qquad  1 - (\hat{\lambda}_{K+1} / \hat{\lambda}_K)  >   t, \\
K+1, &\qquad \mbox{otherwise}.  
\end{cases}
\] 
For the $8$ data sets in Table~\ref{tab:realdata0}, if we choose $t  = 0.1$ as suggested, then $M = K+1$ for the Simmons and Caltech data sets, and $M=K$ for all others.   
The insight is that, if a data set fits with the ``strong signal" profile, 
then we use exactly $K$ eigenvectors for clustering, but if it  
fits with the ``weak signal" profile, we may need to use more than $K$ eigenvectors.   
Our analysis below shows  that the 
Simmons and Caltech data sets fit with the ``weak signal" profile, while all other data sets fit with the ``strong signal" profile.   

We illustrate our points with the scree plot and the Rayleigh  quotient. Let $\ell \in \mathbb{R}^n$ be the true community label vector, and let 
\[
S_k = \{1 \leq i \leq n:  \ell_i = k\}, \qquad 1 \leq k \leq K. 
\]
For any vector $x \in \mathbb{R}^n$,  
define {\it normalized Rayleigh quotient} \cite{fisher36lda}:   
\[
Q(x) =  1 - \frac{\mbox{Within-Class-Variance}}{\mbox{Total Variance}} = \frac{\mbox{Between-Class-Variance}}{\mbox{Total Variance}}, 
\] 
where Total Variance, Within-Class Variance, and Between-Class-variance are 
$\sum_{i = 1}^n (x_i - \bar{x})^2$,  $\sum_{k = 1}^K  \sum_{i \in S_k}  (x_i - \bar{x}_k)^2$,  
and $\sum_{k = 1}^K  (|S_k| \cdot  (\bar{x}_k  - \bar{x})^2)$, respectively ($\bar{x}$ is the overall mean of $x_i$ and $\bar{x}_k$ is the mean of $x_i$ over all $i \in S_k$).  
 Rayleigh quotient is a well-known measure for the clustering utility of $x$. Note that $0 \leq Q(x) \leq 1$ for all $x$,   $Q(x) = 1$ when $x = \ell$, and $Q(x) \approx 0$ when $x$ is a randomly generated vector.

\begin{figure}[tb] 
\centering
\includegraphics[width=.4\textwidth, height = .26\textwidth]{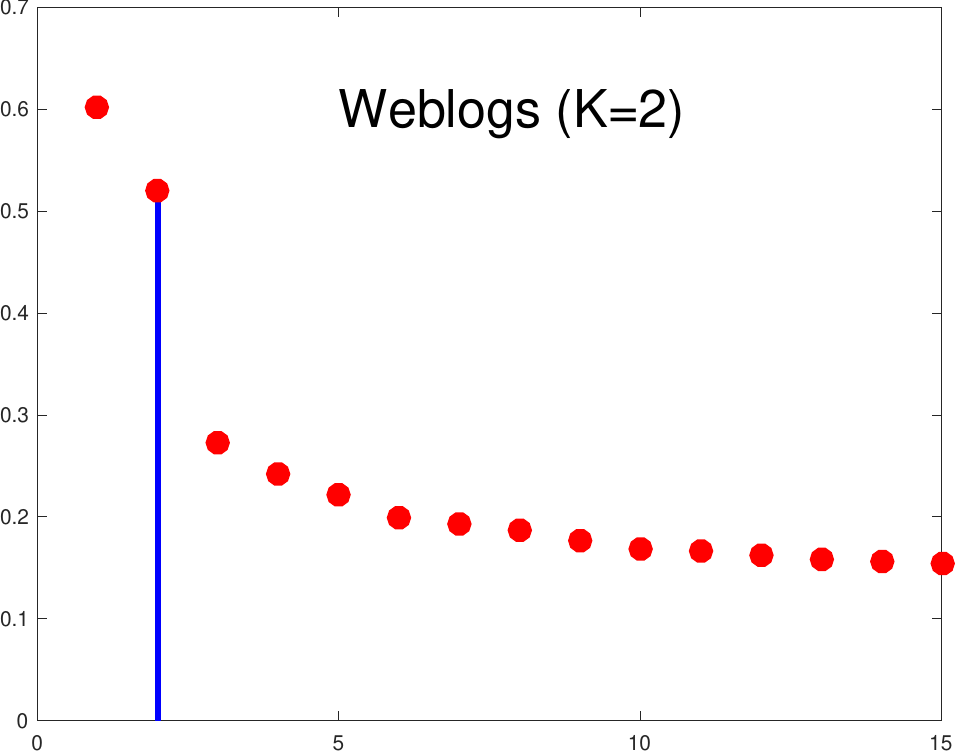}
\includegraphics[width=.4\textwidth, height = .26\textwidth]{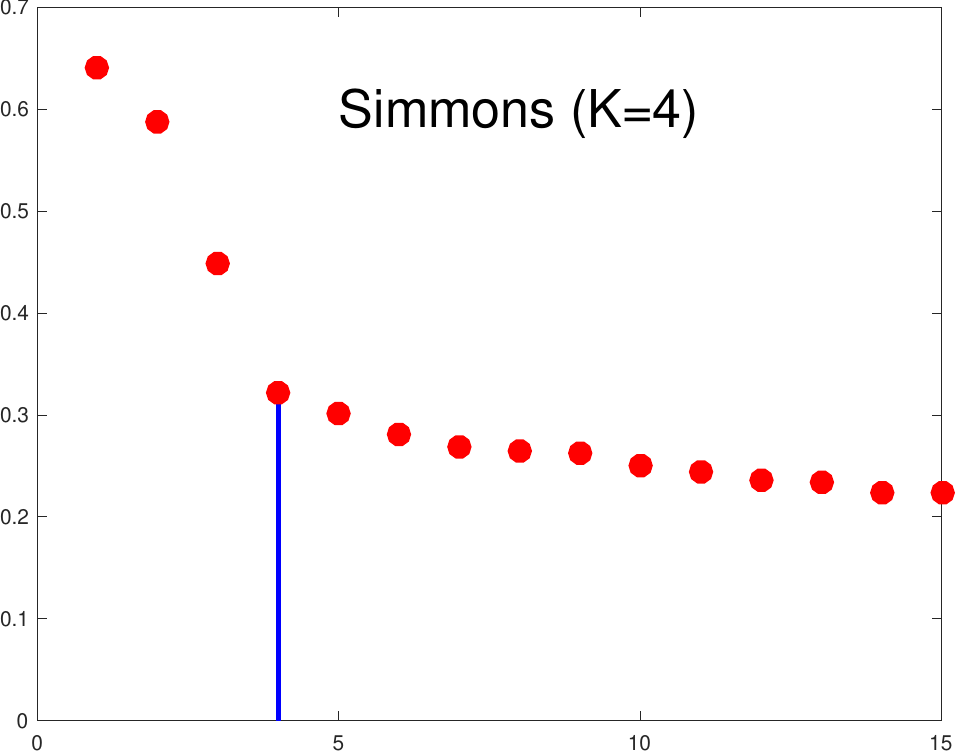}\\
\vspace{1em}
\includegraphics[width=.4\textwidth, height = .26\textwidth]{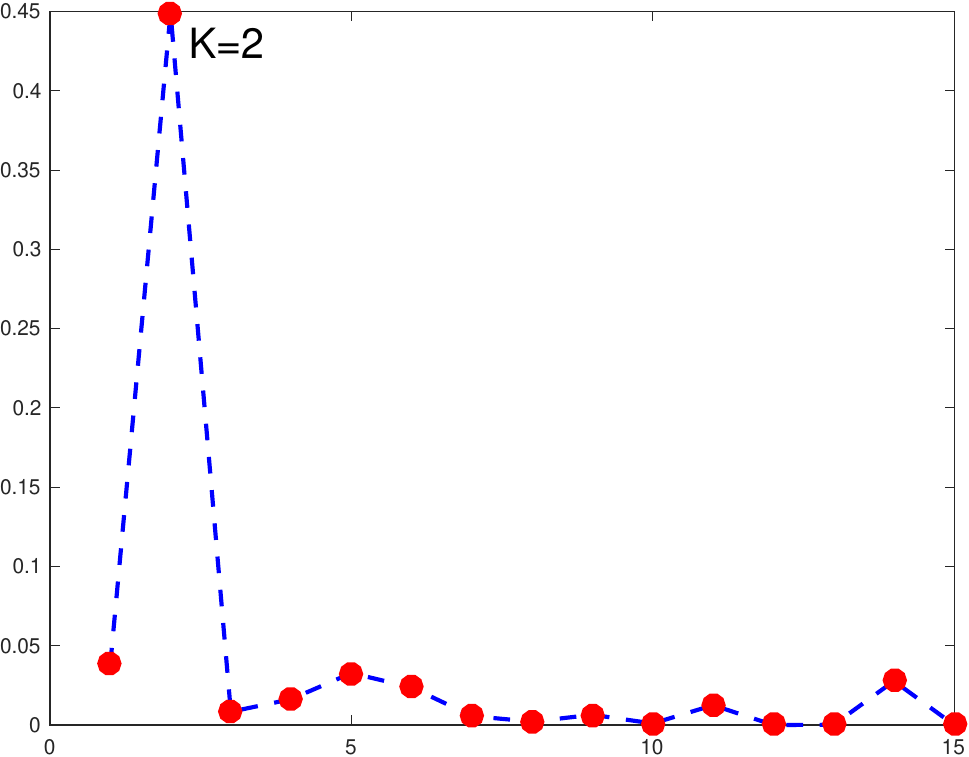}
\includegraphics[width=.4\textwidth, height = .26\textwidth]{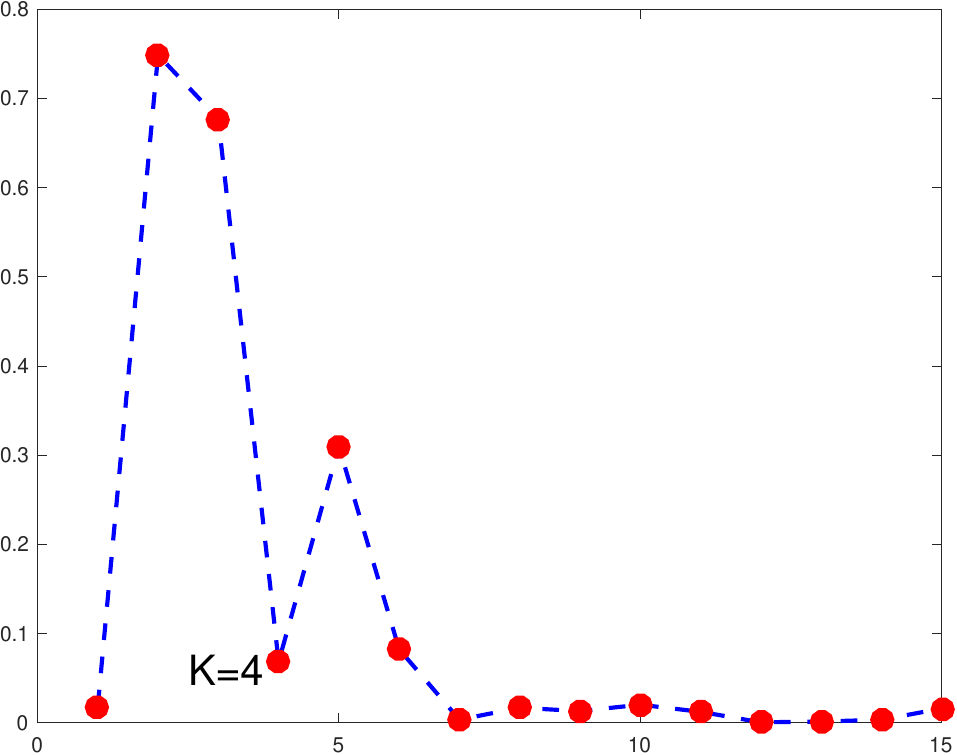}
\caption{A typical ``strong signal" dataset (Weblogs, left panels) and a typical ``weak signal" dataset (Caltech, right panels). The top two figures display the absolute eigenvalues. We observe there is a relatively large gap between $|\hat{\lambda}_K|$ and $|\hat{\lambda}_{K+1}|$ in a ``strong signal" profile and a relatively small gap in a ``weak signal" profile. The bottom two plots display the Rayleigh quotients $Q(\hat{\xi}_k)$. We observe that $Q(\hat{\xi}_k)$ for $k=K+1,K+2,\ldots$ are all small in a ``strong signal" profile but some of  them  are large in a ``weak signal" profile.}  \label{fig:scree1}
\end{figure}

Fix $\delta=0.1$. Let $\hat{\lambda}_1, \hat{\lambda}_2, \ldots, \hat{\lambda}_{K+1}$ be the $(K+1)$ 
eigenvalues of $L_{\delta}$ with largest magnitude and let $\hat{\xi}_1, \hat{\xi}_2, \ldots, \hat{\xi}_{K+1}$ 
be the corresponding eigenvectors.   
Below are some features that help differentiate a ``strong signal" setting from a 
``weak signal" setting.   
\begin{itemize} 
\item In the scree plot,  we expect to see a relatively large gap between $\hat{\lambda}_{K}$ and $\hat{\lambda}_{K+1}$ when the ``signal" is strong, and a relatively small gap  if the  ``signal" is relatively weak.  
\item In a ``strong signal" setting, we expect to see that the Rayleigh  quotient $Q(\hat{\xi}_k)$  is  relatively large for $k = K$, but  is  relatively small for $k = K+1, K+2$, etc. In a ``weak signal" setting,  we may observe that a relatively large Rayleigh quotient  $Q(\hat{\xi}_{k})$ for $k = K+1, K+2$, etc.,   and $Q(\hat{\xi}_K)$ can be relatively small. 
\end{itemize}   
The points are illustrated in Figure \ref{fig:scree1} with the Weblog data and Simmons data, which are believed to be a typical ``strong signal" dataset and a typical ``weak signal" dataset, respectively. 
 We  note that the first eigenvector consists of global information of $L_{\delta}$ and it alone does not have much utility for clustering. Therefore, the corresponding Rayleigh quotient $Q(\hat{\xi}_1)$ is usually small. In SCORE (e.g., \eqref{newSCORE}),  we use $\hat{\xi}_1$ for normalization, but not directly for clustering. 
 
Table~\ref{tab:RQ} shows the Rayleigh quotients of all $8$ datasets. 
We found that the $(K+1)$th eigenvector contains almost no information of community labels, except for Caltech and Simmons. This agrees with our findings that Caltech and Simmons fit with the ``weak signal" profile.

\begin{table}[htb!]  
\centering
\scalebox{.9}{
\begin{tabular}{l |  lll  l  l   ll ll ll ll}
\hline
Dataset &  Polblogs & Karate & Dolphins & Football & Polbooks & UKfaculty & Simmons & Caltech \\
\hline
Eigen(K)&$\bf{0.36}$ & $\bf{0.76}$  & $\bf{0.60}$ &$\bf{0.45}$ &$\bf{0.63}$ & $\bf{0.80}$&    0.04 &   0.25 \\
Eigen(K+1)&0.02 & 0.07   & 0.00  &  0.01 &0.01 &   0.11   & $\bf{0.20}$ &   $\bf{0.47}$ \\%
Eigen(K+2)&0.06 & 0.05   & 0.01 & 0.00&0.01 &   0.00  &  0.13  & 0.06\\%
Eigen(K+3) &0.01 & 0.01 & 0.01 &  0.01 &0.01&    0.00&    0.15 &    0.11 \\%
\hline
Eigen(K)&$\bf{0.45}$ & $\bf{0.81}$  & $\bf{0.79}$ &$\bf{0.48}$ &$\bf{0.79}$ & $\bf{0.89}$&    0.07 &   0.32 \\
Eigen(K+1)&0.02 & 0.02   & 0.00  &  0.22 &0.01 &   0.06   & $\bf{0.31}$ &   $\bf{0.54}$ \\%
Eigen(K+2) &0.03 & 0.01   & 0.00  & 0.00&0.00 &   0.00  &  0.08  & 0.03\\%
Eigen(K+3) &0.01 & 0.01 & 0.02 &  0.02 &0.00&    0.00&    0.00 &    0.09\\%
\hline
\end{tabular}
}
\caption{Rayleigh quotient $Q(\hat{\xi}_k)$ for $8$ networks. The first four rows are for eigenvectors of the adjacency matrix, and the last four rows are for eigenvectors of the regularized graph Laplacian. Except for Simons and Caltech, the $(K+1)$th eigenvector of all other datasets contains almost no information.} \label{tab:RQ} 
\end{table}

How to choose between $K=M$ and $K=M+1$?
The scree plot could potentially be a good way to estimate how much information is contained in each eigenvector. If the $K$th and $(K+1)$th eigenvalues are close,  it is  likely that the $(K+1)$th eigenvector also contains information. To measure ``closeness'', we propose to use the quantity $1 - [\hat{\lambda}_{K+1}/\hat{\lambda}_K]$ with a scale-free tuning parameter $t=0.1$. This seems to work well on  all the 8  datasets. See Table~\ref{tab:tuning}.

\begin{table}[htb!]  
\centering
\scalebox{.9}{
\begin{tabular}{l |  c  | c  }
\hline
Dataset &  Adjacency matrix & Regularized graph Laplacian \\
\hline
Polblogs  &0.5997 & 0.5223   \\
Karate  & 0.4140 & 0.1768  \\
Dolphins &0.1863&0.2027  \\
Football  & 1.9255&0.1414    \\
Polbooks  & 0.5034 &0.2246   \\
UKfaculty &0.3139&0.3336   \\
Simmons  & $\bf{0.0804}$ & $\bf{0.0533}$    \\
Caltech & $\bf{0.0777}$ & $\bf{0.0236}$     \\
\hline
\end{tabular}
}
\caption{The quantity $1 - [\hat{\lambda}_{K+1}/\hat{\lambda}_K]$ for $8$ network  data sets, where $\hat{\lambda}_k$ are from the adjacency matrix (left) and the regularized graph Laplacian (right). With a threshold $t=0.1$, this criterion successfully selects $M  = K +1$ for Simmons and Caltech and $M=K$  for all others.} \label{tab:tuning} 
\end{table}

%%%%%%%%%%%%
%%%%%%%%%%%%
%%%%%%%%%%%%
\section{Proofs} \label{sec:proof}  
We now prove Theorems \ref{thm:Hamming},  Corollary \ref{cor:Hamm}, and 
Theorem \ref{thm:main}.

%%%%%%%%%%%%
%%%%%%%%%%%%
%%%%%%%%%%%%
\subsection{Analysis of empirical eigenvectors}  \label{subsec:EigenAnalysis}
Recall that $\lambda_k$ and $\hat{\lambda}_k$ denote the $k$th largest eigenvalue (in magnitude) of $\Omega$ and $A$, respectively, and $\xi_k$ and $\hat{\xi}_k$ denote the respective eigenvectors. Define
\[
\beta_n = |\lambda_K(G^{1/2}PG^{1/2})|, \qquad \mbox{where}\quad G = K\|\theta\|^{-2}\cdot \mathrm{diag}\bigl(\|\theta^{(1)}\|^2, \ldots,\|\theta^{(K)}\|^2\bigr). 
\]
Let $\|M\|_{2\to\infty}$ denote the maximum row-wise $\ell^2$-norm of a matrix $M$. 
The key technical tool we need in the proof is the following lemma: 
%%%%%%%%%%%%%%%%%%%%
%%%%%%%%%%%%%%%%%%%%
\begin{thm} \label{thm:Aeigenvecs}
Under conditions of Theorem~\ref{thm:Hamming}, write $\hat{\Xi}_0=[\hat{\xi}_2,\hat{\xi}_3,\ldots,\hat{\xi}_K]$, $\Xi_0=[\xi_2,\xi_3,\ldots,\xi_K]$, and $\Lambda_0=\diag(\lambda_2,\lambda_3,\ldots,\lambda_K)$. With probability $1-o(n^{-3})$, there exists an orthogonal matrix $O\in\mathbb{R}^{K-1,K-1}$ (which depends on $A$ and is stochastic) such that 
\[
\|\hat{\Xi}_{0}O - A\Xi_{0}\Lambda_0^{-1}\|_{2\to\infty}\leq \frac{C\theta_{\max}K^5\sqrt{\theta_{\max}\|\theta\|_1\log(n)}}{\beta_n\|\theta\|^3}. 
\]
\end{thm}

Theorem~\ref{thm:Aeigenvecs} is an extension of equation (C.71) in \cite{MSCORE} (this equation appears in the proof of Lemma 2.1 of \cite{MSCORE}).  Lemma 2.1 of \cite{MSCORE} assumes that $|\lambda_2|,|\lambda_3|,\ldots,|\lambda_K|$ are at the same order, but here we allow them to be at different orders. The proof also needs some modification. 

{\bf Remark}. In the bound in Theorem~\ref{thm:Aeigenvecs}, the power of $K$ can be further reduced by adding mild regularity conditions on $|\lambda_2|,|\lambda_3|,\ldots,|\lambda_K|$. For example, if we assume $|\lambda_2|,|\lambda_3|,\ldots,|\lambda_K|$ can be grouped into $s=O(1)$ groups such that $\kappa(I)\leq C$ for each group (see the statement of Lemma~\ref{lem:AbbeFan17new} for the definition of $\kappa(I)$), then the power of $K$ can be reduced from $K^5$ to $K\sqrt{K}$. In fact, the setting in \cite{MSCORE} corresponds to a special case of $s=2$.  

\bigskip
\noindent
{\it Proof of Theorem~\ref{thm:Aeigenvecs}}: Re-arrange the $(K-1)$ eigenvalues in the descending order, i.e., $\lambda_{(2)}\geq \lambda_{(3)}\geq \ldots\geq\lambda_{(K)}$. We first assume that all these eigenvalues are positive and use the following procedure to divide them into groups: 
\begin{itemize}
\item Initialize: $k=1$ and $m=2$. 
\item Compute the eigen-gaps $g_s$, where $g_s= \lambda_{(s)}-\lambda_{(s+1)}$, for $m\leq s\leq K-1$, and $g_K=\lambda_{(K)}$. Let $s^*=\min\{m\leq s\leq K: g_s\geq K^{-1}\lambda_{(m)}\}$. Since $\lambda_{(m)}=\sum_{s=m}^Kg_s$, such $s^*$ must exist.
\item Group $\lambda_{(m)},\lambda_{(m+1)},\ldots,\lambda_{(s^*-1)}, \lambda_{(s^*)}$ together as the $k$th group. 
\item If $s^*=K$, terminate; otherwise, increase $k$ by 1, reset $m=s^*+1$, and repeat the above steps to obtain the next group.  
\end{itemize}
For each group $k$, let $I_k$ be the corresponding index set in the original order, i.e., group $k$ consists of eigenvalues $\lambda_j$ for all $j\in I_k$. Define the eigengap associated with group $k$ as
\beq \label{eigGap}
\delta(I_k) = \min\left\{  \min_{\substack{1\leq i,j\leq K\\ i\in I_k, \, j\notin I_k}} |\lambda_i-\lambda_j|, \;\;\; \min_{i\in I_k}|\lambda_i| \right\}. 
\eeq
The above grouping procedure, as well as the first inequality of condition \eqref{condition1c}, guarantees that
\beq \label{groupProperty}
\max_{j\in I_k}|\lambda_j|\leq K\cdot \delta(I_k), \qquad \mbox{for each group $k$}. 
\eeq
When some of the $(K-1)$ eigenvalues are negative, we first partition these eigenvalues into two subsets, one consists of positive eigenvalues, and the other consists of negative ones. We directly apply the grouping procedure in the first subset. In the second subset, we take absolute values, sort in the descending order, and then apply the above grouping procedure. Finally, we combine the two collections of groups. The resulting groups still satisfy \eqref{groupProperty}. 

We then prove the following technical lemma, which extends Theorem 2.1 of \cite{abbe2017entrywise} and Lemma C.3 of \cite{MSCORE}. 
%%%%%%%%%%%%%%%
%%%%%%%%%%%%%%%
%%%%%%%%%%%
\begin{lemma} \label{lem:AbbeFan17new}
Let $M\in\mathbb{R}^{n,n}$ be a symmetric random matrix, where $\mathbb{E}M=M^*$ for a rank $K_0$ matrix $M^*$. Let $d^*_k$ and $d_k$ be the $k$th largest nonzero eigenvalue of $M^*$ and $M$, respectively, and let $\eta^*_k$ and $\eta_k$ be the corresponding eigenvector, respectively, $1\leq k\leq K_0$. Consider a partition
\[
\{1,2,\ldots,K_0\}= I\cup \bigl(\cup_{k=1}^N I_k), \qquad\mbox{where}\quad I=\{s+1,s+2,\ldots,s+r\},
\]
with $s$ and $r$ being two integers such that $1\leq r\leq K_0$ and $0\leq s\leq K_0-r$, and where each of $I_1,I_2, \ldots,I_N$ contains consecutive indices. For any index subset $B\subset\{1,2,\ldots,K_0\}$, define 
\[
\delta(B)=\min\bigl\{ \min_{i\in B, j\notin B}\{|d^*_i-d^*_j|\}, \;\min_{i\in B}|d^*_i|  \bigr\}, \qquad\mbox{and}\qquad \kappa(B)=\bigl(\max_{i\in B}\{|d_i^*|\}\bigr)/\delta(B).
\]
Let $D=\diag(d_{s+1},\ldots,d_{s+r})$, $D^*=\diag(d^*_{s+1},\ldots,d^*_{s+r})$, 
\[
U=[\eta_{s+1}, \eta_{s+2},\ldots,\eta_{s+r}],\qquad \mbox{and} \qquad U^*=[\eta^*_{s+1}, \eta^*_{s+2},\ldots,\eta^*_{s+r}].
\]
Let $M^*_{m,\cdot}$ denote the $m$-th row of $M^*$, for $1\leq m\leq n$. Suppose for a number $\gamma>0$, the following assumptions are satisfied:
\begin{itemize}
\item A1 (Incoherence): $\max_{1\leq m\leq n}\|M^*_{m,\cdot}\|\leq \gamma\Delta^*$, where $\Delta^*=\delta(I)$. 
\item A2 (Independence): For any $1\leq m\leq n$, the entries of the $m$-th row and column of $M$ are independent with the other entries. 
\item A3 (Spectral norm concentration): For a number $\delta_0\in (0,1)$, $\mathbb{P}(\|M-M^*\|\leq \gamma\Delta^*)\geq 1-\delta_0$.  
\item A4 (Row concentration): There is a number $\delta_1\in (0,1)$ and a continuous non-decreasing function $\varphi(\cdot)$ with $\varphi(0)=0$ and $\varphi(x)/x$ being non-increasing in $\mathbb{R}^+$ such that, for any $1\leq m\leq n$ and non-stochastic matrix $Y\in\mathbb{R}^{n,r}$, 
\[
\mathbb{P}\left(\|(M-M^*)_{m,\cdot}Y\|_2 \leq \Delta^*\|Y\|_{2\to\infty}\varphi\Bigl(\frac{\|Y\|_F}{\sqrt{n}\|Y\|_{2\to\infty}}\Bigr)  \right)\geq 1-\delta_1/n. 
\]   
\end{itemize}
With probability $1-\delta_0-2\delta_1$, for an orthogonal matrix $O\in\mathbb{R}^{r,r}$, 
\begin{align} \label{newentrywise}
& \|UO - MU^*(D^*)^{-1}\|_{2\to\infty}\cr
\leq\; & C\bigl[\kappa(\kappa+\varphi(1))(\gamma+\varphi(\gamma)) + \widetilde{\kappa}\gamma\bigr]\cdot \|\widetilde{U}^*\|_{2\to\infty},
\end{align}
where $\widetilde{U}^*=[\eta_1,\eta_2,\ldots,\eta_{K_0}]$ and $\widetilde{\kappa}=\sum_{1\leq k\leq N}\kappa(I_k)$.  
\end{lemma}

We now prove Lemma~\ref{lem:AbbeFan17new}. The proof is a light modification of the proof of Lemma C.3 of \cite{MSCORE}. Fix $1\leq m\leq n$. Let $M^{(m)}$ be the matrix by setting the $m$th row and the $m$th column of $M$ to be zero. Let $\eta_1^{(m)},\eta_2^{(m)},\ldots,\eta_n^{(m)}$ be the eigenvectors of $M^{(m)}$. Write $U^{(m)}=[\eta_{s+1}^{(m)},\ldots,\eta_{s+r}^{(m)}]$. Let $H=U'U^*$, $H^{(m)}=(U^{(m)})'U^*$ and $V^{(m)}=U^{(m)}H^{(m)}-U^*$. We aim to prove 
\begin{align} \label{lemAbbe-4}
\|M_{m\cdot}V^{(m)}\| \leq &  6(\kappa + \widetilde{\kappa})\gamma \Delta^*\|\widetilde{U}^*\|_{2\to\infty}\cr
 &+ \Delta^*\varphi(\gamma)\bigl(4\kappa \|UH\|_{2\to\infty} + 6\|U^*\|_{2\to\infty}\bigr).
\end{align}
Once \eqref{lemAbbe-4} is obtained, the proof is almost identical to the proof of (B.26) in \cite{abbe2017entrywise}, except that we plug in \eqref{lemAbbe-4} instead of (B.32) in \cite{abbe2017entrywise}. This is straightforward, so we omit it. 
What remains is to prove \eqref{lemAbbe-4}. In the proof of \cite[Lemma 5]{abbe2017entrywise}, it is shown that
\begin{align*}
\|M_{m\cdot}V^{(m)}\| & \leq \|M^*_mV^{(m)}\| + \|(M-M^*)_{m\cdot}V^{(m)}\|, \cr
\|(M-M^*)_{m\cdot}V^{(m)}\| &\leq \Delta^*\varphi(\gamma)\bigl(4\kappa \|UH\|_{2\to\infty} + 6\|U^*\|_{2\to\infty}\bigr).
\end{align*}
Combining them gives
\beq \label{lemAbbe-1}
\|M_{m\cdot}V^{(m)}\|\leq \|M^*_{m\cdot}V^{(m)}\| + \Delta^*\varphi(\gamma)\bigl(4\kappa \|UH\|_{2\to\infty} + 6\|U^*\|_{2\to\infty}\bigr).
\eeq
We further bound the first term in \eqref{lemAbbe-1}. Define
\[
I_0 = \cup_{\{1\leq k\leq N: \, \max_{j\in I_k}|d_j^*|>\max_{j\in I}|d_j^*|\}}I_k. 
\]
In other words, $I_0$ is the union of groups of eigenvalues such that the largest absolute eigenvalue in that group is larger than $\|D^*\|$.  
Let $\widetilde{M}^*=\sum_{j\in I_0}d_j^*\eta_j^*(\eta_j^*)'$. 
\begin{align*} 
\|M^*_{m\cdot}V^{(m)}\|& \leq \|\widetilde{M}_{m\cdot}^*V^{(m)}\| + \|(M^*_{m\cdot}-\widetilde{M}_{m\cdot}^*)V^{(m)}\|\cr
&\leq \|\widetilde{M}_{m\cdot}^*V^{(m)}\| + \|M^*-\widetilde{M}^*\|_{2\to\infty}\|V^{(m)}\|\cr
&\leq   \|\widetilde{M}_{m\cdot}^*V^{(m)}\| + 6\gamma \|M^*-\widetilde{M}^*\|_{2\to\infty},
\end{align*}
where the last line uses $\|V^{(m)}\|\leq 6\gamma$, by (B.12) of \cite{abbe2017entrywise}. Note that
$M^*-\widetilde{M}^*=\sum_{j\notin I_0}d_j^*\eta_j^*(\eta_j^*)'$. By definition of $I_0$, for any $j\notin I_0$, $|d_j^*|\leq \max_{i\in I}|d_i^*|\leq \kappa\Delta^*$. It follows that
\[
\|M^*-\widetilde{M}^*\|_{2\to\infty}\leq \bigl(\max_{j\notin I_0}|d_j^*|\bigr)\|\widetilde{U}^*\|_{2\to\infty}\leq \kappa \Delta^*\|\widetilde{U}^*\|_{2\to\infty}. 
\]
Combining the above gives
\beq \label{lemAbbe-2}
\|M^*_{m\cdot}V^{(m)}\|\leq \|\widetilde{M}_{m\cdot}^*V^{(m)}\| + 6\kappa \gamma \Delta^*\|\widetilde{U}^*\|_{2\to\infty}. 
\eeq 
Without loss of generality, we assume all groups except for $I$ are contained in $I_0$, i.e., $I_0=\cup_{k=1}^NI_k$.  
Let $D_k^*=\mathrm{diag}(d_j^*)_{j\in I_k}$, $U_k^*=[\eta^*_j]_{j\in I_k}$, $U_k=[\eta_j]_{j\in I_k}$, $U_k^{(m)}=[\eta^{(m)}_j]_{j\in I_k}$, and $H_k^{(m)}=(U_k^{(m)})'U_k^*$. Then, 
\[
\widetilde{M}^*=\sum_{k=1}^N U_k^*\Lambda^*_k(U_k^*)'. 
\]
Similar to (B.12) of \cite{abbe2017entrywise}, we have $\|U_k^{(m)}H_k^{(m)}-U^*_k\|\leq 6\gamma_k$, where $\gamma_k$ is defined in the same way as $\gamma$ but is with respect to the eigen-gap of group $k$, which is $\Delta^*_k\equiv \delta(I_k)$. It is not hard to see that $\gamma_k=\gamma\Delta^*/\Delta^*_k$. Therefore, 
\beq \label{temp}
\|U_k^{(m)}H_k^{(m)}-U^*_k\|\leq 6\gamma\Delta^*/\Delta^*_k, \qquad 1\leq k\leq N.  
\eeq
By mutual orthogonality of eigenvectors, $(U_k^{(m)})'U^{(m)}=0$, and $(U^*_k)'U^*=0$. Additionally, we have $\|U_k^{(m)}\|=1$ and $\|H_k^{(m)}\|\leq 1$. It follows that
\begin{align*}
\|\widetilde{M}_{m\cdot}^*V^{(m)}\| &\leq \sum_{k=1}^N \|e_m'[U_k^*\Lambda^*_k(U_k^*)'][U^{(m)}H^{(m)}-U^*]\|\cr
&= \sum_{k=1}^N \bigl\|e_m'[U_k^*\Lambda^*_k(U_k^*)']U^{(m)}H^{(m)}\bigr\|  \quad \mbox{(by mutual orthogonality)}\cr
&\leq \sum_{k=1}^N \bigl\|e_m'[U_k^*\Lambda^*_k(U_k^*)']U^{(m)}\bigr\|\cr
&= \sum_{k=1}^N \bigl\|e_m'U_k^*\Lambda^*_k(U_k^* - U_k^{(m)}H_k^{(m)})'U^{(m)}\bigr\| \quad \mbox{(by mutual orthogonality)} \cr
&\leq \sum_{k=1}^N \bigl\|e_m'U_k^*\Lambda^*_k(U_k^* - U_k^{(m)}H_k^{(m)})'\bigr\|\cr
&\leq \sum_{k=1}^N \|U_k^*\|_{2\to\infty}\cdot\|\Lambda_k^*\|\cdot \|U_k^* - U_k^{(m)}H_k^{(m)}\| \cr
&\leq \sum_{k=1}^N 6(\|\Lambda_k\|^*/\Delta^*_k)\cdot \gamma\Delta^* \|U_k^*\|_{2\to\infty} \qquad \mbox{(by \eqref{temp})}\cr
&\leq 6\gamma\Delta^*\cdot \bigl(\sum_{k=1}^N\kappa(I_k)\bigr)\cdot \|\widetilde{U}^*\|_{2\to\infty}  \quad \mbox{(note that $\|U_k^*\|_{2\to\infty}\leq\| \widetilde{U}^*\|_{2\to\infty}$).}
\end{align*} 
We plug it into \eqref{lemAbbe-2} and use the definition of $\widetilde{\kappa}$. It gives
\beq \label{lemAbbe-3}
\|M^*_{m\cdot}V^{(m)}\|\leq 6(\kappa + \widetilde{\kappa})\gamma \Delta^*\|\widetilde{U}^*\|_{2\to\infty}. 
\eeq
Combining \eqref{lemAbbe-3} with \eqref{lemAbbe-1} gives \eqref{lemAbbe-4}. Then, the proof of Lemma~\ref{lem:AbbeFan17new} is complete. 

\bigskip

We now apply Lemma~\ref{lem:AbbeFan17new} to prove the claim. For the groups in \eqref{groupProperty}, they satisfy that
\[
\kappa(I_k)\leq K,\qquad \mbox{and}\qquad \delta(I_k)\geq K^{-1}|\lambda_K|\geq K^{-2}\beta_n\|\theta\|^2. 
\]
We fix $I$ to be one of the groups, and let $I_1,\ldots,I_N$ be the remaining groups. We apply Lemma~\ref{lem:AbbeFan17new} to $M=A$, $M^*=\Omega=\mathrm{diag}(\Omega)+(A-\mathbb{E}A)$, and  
\[
\Delta^*\asymp \beta_nK^{-2}\|\theta\|^2, \qquad \gamma\asymp \frac{K^2\sqrt{\theta_{\max}\|\theta\|_1}}{\beta_n\|\theta\|^2}.
\]
We construct $\varphi(\gamma)$ in the same way as in Lemma C.3 of \cite{MSCORE}. It satisfies that $\varphi(\gamma)\leq C\gamma\sqrt{\log(n)}$. Similarly as in the proof of Lemma C.3, we can show that conditions A1-A4 are satisfied. Write 
\[
\hat{\Xi}_{01}=[\hat{\xi}_i]_{i\in I}, \qquad \Xi_{01}=[\xi_i]_{i\in I}, \qquad \mbox{and}\quad \Lambda_1=\diag(\lambda_i)_{i\in I}. 
\]
It follows from \eqref{newentrywise} that there exists an orthogonal matrix $O\in\mathbb{R}^{|I|\times|I|}$ such that 
\[
 \|\hat{\Xi}_{01}O - A\Xi_{01}\Lambda_1^{-1}\|_{2\to\infty}
\leq C\frac{K^4 \sqrt{\theta_{\max}\|\theta\|_1\log(n)}}{\beta_n\|\theta\|^2}\|\Xi\|_{2\to\infty}. 
\]
By Lemma B.2 of \cite{MSCORE}, $\|\Xi\|_{2\to\infty}=O(\sqrt{K}\|\theta\|^{-1}\theta_{\max})$. Plugging it into the above inequality, we find that
\beq \label{entrywise-2}
\|\hat{\Xi}_{01}O - A\Xi_{01}\Lambda_1^{-1}\|_{2\to\infty}\leq \frac{C\theta_{\max}K^4\sqrt{K\theta_{\max}\|\theta\|_1\log(n)}}{\beta_n\|\theta\|^3}. 
\eeq
The above inequality holds for each group. Note that $\hat{\Xi}_0$ is obtained by putting such $\hat{\Xi}_{01}$ together. When $B=[B_1,B_2,\ldots,B_N]$, it holds that $\|B\|_{2\to\infty}\leq \sqrt{\sum_k\|B_k\|^2_{2\to\infty}}\leq \sqrt{K}\max_{k}\|B_k\|_{2\to\infty}$. Combining it with \eqref{entrywise-2} gives the claim. \qed

\subsection{Proof of Theorem~\ref{thm:Hamming}}
The rationale of SCORE guarantees that the rows of $R$ take only $K$ distinct values $v_1,v_2,\ldots,v_K\in\mathbb{R}^{K-1}$. 
Below, we first derive a crude high-probability bound for $\mathrm{Hamm}(\widehat{\Pi},\Pi)$ without using Theorem~\ref{thm:Aeigenvecs}. This bound implies that each k-means center is close to one of the true $v_k$. Next, we use Theorem~\ref{thm:Aeigenvecs} to derive a sharper bound for $\mathbb{E}[\mathrm{Hamm}(\widehat{\Pi},\Pi)]$. 

We start from deriving a crude bound for  $\mathrm{Hamm}(\widehat{\Pi},\Pi)$. Let $\beta_n$ be the same as in Section~\ref{subsec:EigenAnalysis}. By Lemma B.1 of \cite{MSCORE}, $C^{-1}K^{-1}\|\theta\|^2\leq \lambda_1\leq \|\theta\|^2$, and $|\lambda_K|\asymp K^{-1}\beta_n\|\theta\|^2$. It follows that  
\[
C^{-1}\sqrt{K}(|\lambda_K|/\sqrt{\lambda_1})\leq \beta_n\|\theta\|\leq CK(|\lambda_K|/\sqrt{\lambda}_1). 
\]
Therefore, the assumption $s_n\geq a_1^{-1} K^4\sqrt{\log(n)}$ guarantees that 
\beq \label{thm-Hamm-const}
\frac{\theta_{\max}K^4\sqrt{K\theta_{\max}\|\theta\|_1\log(n)}}{\beta_n\theta_{\min}\|\theta\|^2}\leq a_1.  
\eeq
Let $O$ be the orthogonal matrix in Theorem~\ref{thm:Aeigenvecs}. By Lemma 2.1 of \cite{MSCORE}, with probability $1-o(n^{-3})$, there exists $\omega\in \{\pm 1\}$ such that  
\beq \label{thm-Hamm-1}
\|\omega \hat{\xi}_1-\xi_1\|_\infty\leq \frac{C\theta_{\max}K\sqrt{\theta_{\max}\|\theta\|_1\log(n)}}{\|\theta\|^{3}}, \qquad \|\hat{\Xi}_0O- \Xi_0\|_F \leq  C\frac{K\sqrt{K\theta_{\max} \|\theta\|_1}}{\beta_n\|\theta\|^2}. 
\eeq
By Lemma B.2 of \cite{MSCORE}, $\xi_1(i)\geq C^{-1}\theta_i/\|\theta\|\geq C^{-1}\theta_{\min}/\|\theta\|$. By choosing $a_1$ appropriately small, the condition on $s_n$ guarantees that $\|\hat{\xi}_1-\xi_1\|_\infty\leq \xi_1(i)/3$, for any $1\leq i\leq n$. Then, we can use a proof similar to that in Lemma C.5 of \cite{MSCORE} to show that, with probability $1-o(n^{-3})$, there exists an orthogonal matrix $H$ such that 
\[
\sum_{i=1}^n\|H\hat{r}_i-r_i\|^2\leq \frac{\|\hat{\Xi}_0O- \Xi_0\|^2_F}{(\min_{1\leq i\leq n}\theta_i/\|\theta\|)^2 }\leq \frac{CK^3\theta_{\max}\|\theta\|_1}{\theta_{\min}^2\beta_n^2 \|\theta\|^2}. 
\]
Since $\theta_{\max}^2\geq \|\theta\|^2/n$, we can further write that
\beq\label{thm-Hamm-3}
\sum_{i=1}^n\|H\hat{r}_i-r_i\|^2\leq \frac{CnK^3\theta^3_{\max}\|\theta\|_1}{\theta_{\min}^2\beta_n^2 \|\theta\|^4}\leq Cn\cdot \frac{a_1^2}{K^6\log(n)}, 
\eeq
where the last inequality is from \eqref{thm-Hamm-const}. Recall that the rows of $R$ take only $K$ distinct values $v_1,\ldots,v_K$. By Lemma B.3 of \cite{MSCORE}, there exists a constant $c_0>0$ such that, for all $1\leq k\neq \ell\leq K$,  
\beq \label{thm-Hamm-4}
\|v_k-v_\ell\|\geq c_0\sqrt{K}\qquad \mbox{and}\qquad \|v_k\|\leq C\sqrt{K}.  
\eeq
Furthermore, in the proof of Theorem 2.2 of \cite{SCORE}, it was shown that the k-means solution satisfies that 
\[
\mathrm{Hamm}(\widehat{\Pi},\Pi) \leq (3/\delta)^2 \sum_{i=1}^n\|H\hat{r}_i-r_i\|^2,  
\]
where $\delta$ is the minimum distance between two distinct rows of $R$. 
Combining the above gives
\beq \label{thm-Hamm-5}
\mathrm{Hamm}(\widehat{\Pi},\Pi)\leq Ca_1^2 \cdot \frac{n}{K^7 \log(n)}, 
\eeq
where $C$ is a constant that does not depend on $a_1$. 

This crude bound \eqref{thm-Hamm-5} is enough for studying the k-means centers. 
By \eqref{thm-Hamm-5}, the total number of misclustered nodes is $O(n/[K^7\log(n)])$. Also, Condition \eqref{condition1b} implies that each true cluster has at least $c^{-1}_2K^{-1}n$ nodes. This means that each cluster has only a negligible fraction of misclustered nodes. Particularly, each true cluster ${\cal C}_k$ is associated with one and only one k-means cluster, which we denote by $\hat{\cal C}_k$; furthermore, we have $|\hat{\cal C}_k\backslash {\cal C}_k|= O(n/[K^7\log(n)])$ and $|\hat{\cal C}_k\backslash {\cal C}_k| =O(n/[K^7\log(n)])$. The cluster center $\hat{v}_k$ of the cluster $\hat{\cal C}_k$ satisfies that
\[
\hat{v}_k=\frac{1}{|\hat{\cal C}_k|}\sum_{i\in \hat{\cal C}_k}\hat{r}_i. 
\]
Note that $r_i=v_k$ for $i\in {\cal C}_k$. It follows that 
\[
\|H\hat{v}_k-v_k\| \leq \frac{1}{|\hat{\cal C}_k|}\biggl\|  \sum_{i\in \hat{\cal C}_k}(H\hat{r}_i-r_i) \biggr\|+  \frac{1}{|\hat{\cal C}_k|}\biggl\|  \sum_{i\in \hat{\cal C}_k\backslash {\cal C}_k}(r_i-v_k) \biggr\|. 
\]
By \eqref{thm-Hamm-4}, $\|r_i-v_k\|=O(\sqrt{K})$. Furthermore, $|\hat{\cal C}_k|\gtrsim c^{-1}_2K^{-1}n$ and $|\hat{\cal C}_k\backslash {\cal C}_k|=O(n/[K^7\log(n)])$. Combing them with the Cauchy-Schwarz inequality, we find that  
\begin{align*} 
\|H\hat{v}_k-v_k\| &\leq \frac{1}{\sqrt{|\hat{\cal C}_k|}} \cdot \sqrt{\sum_{i\in \hat{\cal C}_k} \|H\hat{r}_i-r_i\|^2}+\frac{|\hat{\cal C}_k\backslash {\cal C}_k|}{c^{-1}_2K^{-1}n}\cdot O(\sqrt{K}) \cr
&\leq \frac{1}{\sqrt{c^{-1}_2K^{-1}n}}\cdot O\Bigl(\sqrt{\frac{n}{K^6\log(n)}}\Bigr) + O\Bigl(\frac{1}{K^5\sqrt{K}\log(n)}\Bigr)\cr
&= O\Bigl(\frac{1}{K^2\sqrt{K\log(n)}}\Bigr). 
\end{align*}
The right hand side is $o(\sqrt{K})$. Let $c_0$ be the same as in \eqref{thm-Hamm-4}. Then, for sufficiently large $n$,
\beq \label{thm-Hamm-6}
\|H\hat{v}_k-v_k\|\leq c_0\sqrt{K}/8, \qquad \mbox{for all }1\leq k\leq K. 
\eeq

Next, we use Theorem~\ref{thm:Aeigenvecs} to get the desired bound for $\mathbb{E}[\mathrm{Hamm}(\widehat{\Pi},\Pi)]$. Let $D$ be the event that \eqref{thm-Hamm-6} holds. We have shown that $\mathbb{P}(D^c)=o(n^{-3})$. It follows that 
\beq \label{thm-Hamm-7}
\mathbb{E}[\mathrm{Hamm}(\widehat{\Pi},\Pi)]=\frac{1}{n}\sum_{i=1}^n\mathbb{P}(\hat{\pi}_i\neq \pi_i)\leq \frac{1}{n}\sum_{i=1}^n\mathbb{P}(\hat{\pi}_i\neq \pi_i, D) + o(n^{-3}). 
\eeq
It remains to bound the probability of making a clustering error on $i$, when the event $D$ holds. Suppose $i\in {\cal C}_k$. On the event $D$, if $\|H\hat{r}_i-r_i\|\leq c_0\sqrt{K}/4$, then 
\[
\|H\hat{r}_i-H\hat{v}_k\|\leq c_0\sqrt{K}/4+c_0\sqrt{K}/8\leq 3c_0\sqrt{K}/8, 
\]
while for any $\ell\neq k$, 
\[
\|H\hat{r}_i-H\hat{v}_\ell\|\geq \|v_k-v_\ell\|-c_0\sqrt{K}/4-c_0\sqrt{K}/8\geq 5c_0\sqrt{K}/8. 
\]
Then, node $i$ must be clustered into $\hat{\cal C}_k$, i.e., there is no error on $i$. This implies that
\beq \label{thm-Hamm-8}
\mathbb{P}(\hat{\pi}_i\neq \pi_i, D)\leq \mathbb{P}\bigl( \|H\hat{r}_i-r_i\|> c_0\sqrt{K}/4 \bigr). 
\eeq

We further study the right hand side of \eqref{thm-Hamm-8}. 
Let $(\hat{\xi}_1,\hat{\Xi}_0, \omega, O)$ be the same as in \eqref{thm-Hamm-1}. Fix $i$. Let $\hat{\Xi}_{0,i}'\in\mathbb{R}^{K-1}$ and $\Xi_{0,i}'\in\mathbb{R}^{K-1}$ denote the $i$th row of $\hat{\Xi}_0$ and $\Xi_0$, respectively. Then,
\[
H\hat{r}_i = \frac{1}{\omega\hat{\xi}_1(i)}O'\hat{\Xi}_{0,i}, \qquad r_i = \frac{1}{\xi_1(i)}\Xi_{0,i}. 
\]
It is seen that 
\[
H\hat{r}_i-r_i =\frac{1}{\omega\hat{\xi}_1(i)}(O'\hat{\Xi}_{0,i}-\Xi_{0,i})-\frac{\omega\hat{\xi}_1(i)-\xi_1(i)}{\omega\hat{\xi}_1(i)}r_i. 
\]
By Lemma B.2 of \cite{MSCORE}, $\xi_1(i)\geq C^{-1}\theta_i/\|\theta\|$. Combining it with \eqref{thm-Hamm-1} gives $\|\omega\hat{\xi}_1-\xi_1\|_\infty=o(\xi_1(i))$. It follows that $\omega\hat{\xi}_1(i)\geq \xi_1(i)/2\geq C^{-1}\theta_i/\|\theta\|$. Additionally, $\|r_i\|\leq C\sqrt{K}$, by \eqref{thm-Hamm-4}. Therefore, with probability $1-o(n^{-3})$,
\begin{align*}
\| H\hat{r}_i-r_i \| &\leq \frac{C\|\theta\|}{\theta_i}\bigl( \|O'\hat{\Xi}_{0,i}-\Xi_{0,i}\| + \sqrt{K}|\omega\hat{\xi}_1(i)-\xi_1(i)|\bigr)\cr
&\leq \frac{C\|\theta\|}{\theta_i}\bigl( \|O'\hat{\Xi}_{0,i}-\Lambda_0^{-1}\Xi_0'A_{\cdot,i}\|  + \|\Lambda_0^{-1}\Xi_0'A_{\cdot,i}-  \Xi_{0,i}\| + \sqrt{K}|\omega\hat{\xi}_1(i)-\xi_1(i)|\bigr)\cr
&\leq \frac{C\|\theta\|}{\theta_{\min}}\bigl( \|\hat{\Xi}_{0}O-A\Xi_0\Lambda_0^{-1}\|_{2\to\infty}+\sqrt{K}\|\omega\hat{\xi}_1-\xi_1\|_\infty\bigr) + \frac{C\|\theta\|}{\theta_i} \|\Lambda_0^{-1}\Xi_0'A_{\cdot,i}-  \Xi_{0,i}\|. 
\end{align*}
We plug in Theorem~\ref{thm:Aeigenvecs} and the first inequality of \eqref{thm-Hamm-1}. It yields 
\begin{align*}
\| H\hat{r}_i-r_i \| & \leq \frac{C\theta_{\max}K^5\sqrt{\theta_{\max}\|\theta\|_1\log(n)}}{\beta_n\theta_{\min}\|\theta\|^2} + \frac{C\|\theta\|}{\theta_i} \|\Lambda_0^{-1}\Xi_0'A_{\cdot,i}-  \Xi_{0,i}\|\cr
&\leq C\sqrt{K}\cdot a_1 + \frac{C_1 \|\theta\|}{\theta_i} \|\Lambda_0^{-1}\Xi_0'A_{\cdot,i}-  \Xi_{0,i}\|,
\end{align*}
where the second inequality is from \eqref{thm-Hamm-const} and the constant $C_1$ does not depend on $a_1$. By choosing an appropriately small $a_1$, we can make the first term $\leq c_0\sqrt{K}/8$. It follows that 
\beq \label{thm-Hamm-9}
\mathbb{P}(\hat{\pi}_i\neq \pi_i, D)\leq \mathbb{P}\left( \frac{C_1\|\theta\|}{\theta_i} \|\Lambda_0^{-1}\Xi_0'A_{\cdot,i}-  \Xi_{0,i}\|>c_0\sqrt{K}/8 \right) + o(n^{-3}). 
\eeq
Note that $A = \Omega + W-\diag(\Omega)$, where $W=A-\mathbb{E}A$ and $\Omega=\Theta\Pi P\Pi'\Theta=\Xi\Lambda\Xi'$. In particular,
\[
\Lambda_0^{-1}\Xi_0'\Omega = \Xi_0'.  
\]
It follows that
\[
\Lambda_0^{-1}\Xi_0'A_{\cdot,i}=\Lambda_0^{-1}\Xi_0'[\Omega+W-\diag(\Omega)]_{\cdot,i} =\Xi_{0,i}+ \Lambda_0^{-1}\Xi_0'W_{\cdot,i} - \Omega(i,i)   \Lambda_0^{-1}\Xi_{0,i}. 
\]
Note that $\Omega(i,i)\leq \theta_i^2$. Additionally, $\|\Lambda_0^{-1}\|= |\lambda_K|^{-1}\asymp K\beta^{-1}_n\|\theta\|^{-2}$ and $\|\Xi_{0,i}\|\leq \|\Xi_0\|\leq 1$. It follows that 
\begin{align*}
\frac{C_1\|\theta\|}{\theta_i} \|\Lambda_0^{-1}\Xi_0'A_{\cdot,i}-  \Xi_{0,i}\| & \leq \frac{C_1\|\theta\|\|\Lambda_0^{-1}\|}{\theta_i}\bigl( \|\Xi_0'W_{\cdot,i}\| + \theta_i^2\bigr)\cr
&\leq \frac{C_2 K}{\theta_i\beta_n\|\theta\|} \|\Xi_0'W_{\cdot,i}\| + \frac{C_2 K\theta_{\max}}{\beta_n\|\theta\|}, 
\end{align*}
where $C_2>0$ is a constant that does not depend on $a_1$. 
The second term is $O(K\beta_n^{-1}\|\theta\|^{-1})$. At the same time, the left hand side of \eqref{thm-Hamm-const} is lower bounded by $K^4\sqrt{K\log(n)}/(\beta_n\|\theta\|)$. Therefore, \eqref{thm-Hamm-const} implies that the second term is $O(1/[K^3\sqrt{K\log(n)}])=o(\sqrt{K})$. Particularly, for sufficiently large $n$, the second term is $\leq c_0\sqrt{K}/16$, i.e.,
\[
\frac{C_1\|\theta\|}{\theta_i} \|\Lambda_0^{-1}\Xi_0'A_{\cdot,i}-  \Xi_{0,i}\|\leq \frac{C_2K}{\theta_i\beta_n\|\theta\|} \|\Xi_0'W_{\cdot,i}\| + (c_0/16)\sqrt{K}.  
\]
We plug it into \eqref{thm-Hamm-9} to get
\begin{align} \label{thm-Hamm-10}
\mathbb{P}(\hat{\pi}_i\neq \pi_i, D)&\leq \mathbb{P}\left( \frac{C_2K}{\theta_i\beta_n\|\theta\|} \|\Xi_0'W_{\cdot,i}\| >(c_0/16) \sqrt{K} \right) + o(n^{-3})\cr
&= \mathbb{P}\left( \|\Xi_0'W_{\cdot,i}\|^2 >\frac{c^2_0}{16^2C_1^2} \cdot \frac{\theta^2_i\beta^2_n\|\theta\|^2}{K} \right) + o(n^{-3})\cr
&= \mathbb{P}\left( \sum_{k=2}^K (e_i'W\xi_k)^2 >\frac{c^2_0}{16^2C_1^2} \cdot \frac{\theta^2_i\beta^2_n\|\theta\|^2}{K} \right) + o(n^{-3}) \cr
&\leq \sum_{k=2}^K \mathbb{P}\left( |e_i'W\xi_k| >\frac{c_0}{16C_1} \cdot \frac{\theta_i\beta_n\|\theta\|}{K} \right) + o(n^{-3}), 
\end{align}
where the last inequality is because of the probability union bound. 

It remains to get a large deviation inequality for $|e_i'W\xi_k|$. Note that
\[
e_i'W\xi_k =\sum_{1\leq j\leq n: j\neq i} \xi_k(j)W(i,j). 
\]
The summands are independent, and $|\xi_k(j)W(i,j)|\leq |\xi_k(j)| \leq C\sqrt{K}\theta_j/\|\theta\|\leq C\sqrt{K}\theta_{\max}/\|\theta\|$ (the bound of $|\xi_k(j)|$ is from Lemma B.2 of \cite{MSCORE}). We shall apply Bernstein's inequality. Note that $\sum_{j}\xi_k^2(j)\mathrm{Var}(W(i,j))\leq \sum_j\xi_k^2(j)\|P\|_{\max}\theta_i\theta_j\leq C\sum_j (K\theta^2_j/\|\theta\|^2)\theta_i\theta_j\leq \theta_i\cdot CK\|\theta\|_3^3/\|\theta\|^2$. It follows from Bernstein's inequality that
\[
\mathbb{P}\bigl( |e_i'W\xi_k|>t\bigr)\leq 2\exp\left( - \frac{t^2/2}{\theta_i\cdot CK\|\theta\|_3^3/\|\theta\|^2 + (t/3)\cdot C\sqrt{K}\theta_{\max}/\|\theta\|} \right), \qquad\mbox{for all }t>0. 
\]
We plug in $t=(c_0/16C_1)\cdot K^{-1}\theta_i\beta_n\|\theta\|$. It follows that
\begin{align} \label{thm-Hamm-11}
\mathbb{P}\biggl( |e_i'W\xi_k|>\frac{c_0}{16C_1} \cdot & \frac{\theta_i\beta_n\|\theta\|}{K} \biggr) \leq 2\exp\left( - \frac{K^{-2}\theta_i^2\beta_n^2\|\theta\|^2}{C_3\theta_i\cdot K \|\theta\|_3^3/\|\theta\|^2 + C_4\theta_i\cdot K^{-1/2}\theta_{\max}\beta_n} \right)\cr
&\leq 2\exp\left( - a_2 \theta_i\cdot \min\left\{ \frac{\beta_n^2\|\theta\|^4}{K^3\|\theta\|_3^3},\; \frac{\beta_n\|\theta\|^2}{K\sqrt{K}\theta_{\max}}  \right\}\right)\cr
&\leq 2\exp\left( - a_2 \theta_i\cdot \min\left\{ \frac{(|\lambda_K|/\sqrt{\lambda_1})^2\|\theta\|^2}{K^2\|\theta\|_3^3},\; \frac{(|\lambda_K|/\sqrt{\lambda_1})\|\theta\|}{K\theta_{\max}}  \right\}\right). 
\end{align}
where $C_3, C_4$ are constants that depend on $(c_0, C_2, C)$, $a_2=\min\{C_3,C_4\}$, and the last inequality is due to $\beta_n\geq \sqrt{K}(|\lambda_K|/\sqrt{\lambda_1})$.  We plug it into \eqref{thm-Hamm-10} to get
\[
\mathbb{P}(\hat{\pi}_i\neq \pi_i, D)\leq 2K\exp\left( - a_2 \theta_i\cdot \min\left\{ \frac{(|\lambda_K|/\sqrt{\lambda_1})^2\|\theta\|^2}{K^2\|\theta\|_3^3},\; \frac{(|\lambda_K|/\sqrt{\lambda_1})\|\theta\|}{K\theta_{\max}}  \right\}\right)+o(n^{-3}). 
\] 
Combining it with \eqref{thm-Hamm-7} gives the desired claim. \qed

\subsection{Proof of Corollary~\ref{cor:Hamm}}
We use an intermediate result in the proof of Theorem~\ref{thm:Hamming}, which is the second last line of \eqref{thm-Hamm-11}. We plug it into \eqref{thm-Hamm-10} and \eqref{thm-Hamm-7} to get
\beq \label{cor-0}
\mathbb{E}[\mathrm{Hamm}(\widehat{\Pi},\Pi)]\leq 2K\sum_{k=1}^K \exp\left( - a_2 \theta_i\cdot \min\left\{ \frac{\beta_n^2\|\theta\|^4}{K^3\|\theta\|_3^3},\; \frac{\beta_n\|\theta\|^2}{K\sqrt{K}\theta_{\max}}  \right\}\right), 
\eeq
where $\beta_n=|\lambda_K(G^{1/2}PG^{1/2})|$, with $G=K\|\theta\|^{-2}\diag(\|\theta^{(1)}\|^2,\ldots,\|\theta^{(K)}\|^2)$.  In this example, from the way $\pi_i$ is generated, by elementary probability, $\|G-I_K\|=O(\sqrt{\log(K)/n})$; moreover, the first eigenvalue of $P$ is $(1-b)+Kb$, and other eigenvalues are all equal to $(1-b)$. It follows that
\[
\beta_n\asymp 1-b. 
\]
Additionally, since $\theta_{\max}\leq C\theta_{\min}$, we have $\|\theta\|_3^3\asymp \|\theta\|^2\bar{\theta}$. It follows that 
\[
\frac{\beta_n^2\|\theta\|^4}{K^3\|\theta\|_3^3}\asymp \frac{n}{\bar{\theta}} \frac{(1-b)^2\|\theta\|^2}{K^3}, \qquad \frac{\beta_n\|\theta\|^2}{K\sqrt{K}\theta_{\max}} \asymp \frac{1}{\bar{\theta}} \frac{(1-b)\|\theta\|}{K\sqrt{K}}. 
\]
The first term is dominating. We plug it into \eqref{cor-0}. The claim follows immediately. \qed

\subsection{Proof of Theorem~\ref{thm:main}}
We have shown in \eqref{thm-Hamm-8} that there is an event $D$ such that $\mathbb{P}(D^c)=o(n^{-3})$ and that on the event $D$,  
\[
\|H\hat{r}_i-r_i\|\leq c_0\sqrt{K}/4 \qquad \Longrightarrow\qquad \hat{\pi}_i=\pi_i. 
\]
Therefore, it suffices to show that, with probability $1-o(n^{-3})$,
\beq \label{thm-main-1}
\|H\hat{r}_i-r_i\|\leq c_0\sqrt{K}/4. 
\eeq
In the equation above \eqref{thm-Hamm-9} and the equation above \eqref{thm-Hamm-10}, we have shown that, as long as $a_1$ in Theorem~\ref{thm:Hamming} is properly small, 
\begin{align} \label{thm-main-4}
\| H\hat{r}_i-r_i \| & \leq  (c_0/8)\sqrt{K} + \frac{C_1\|\theta\|}{\theta_i} \|\Lambda_0^{-1}\Xi_0'A_{\cdot,i}-  \Xi_{0,i}\|\cr
&\leq (c_0/8)\sqrt{K}+ \frac{C_2K}{\theta_i\beta_n\|\theta\|} \|\Xi_0'W_{\cdot,i}\| + (c_0/16)\sqrt{K}\cr
&\leq (3c_0/16) \sqrt{K} + \frac{C_2K}{\theta_i\beta_n\|\theta\|} \sqrt{\sum_{k=2}^K (e_i'W\xi_k)^2}. 
\end{align}
We then apply \eqref{thm-Hamm-11}. In order for the exponent of the right hand side of \eqref{thm-Hamm-11} to be at the order of $\log(n)$, we need
\beq \label{thm-main-2}
\theta_{\min}\cdot \frac{(|\lambda_K|/\sqrt{\lambda_1})^2\|\theta\|^2}{K^2\|\theta\|_3^3}\geq C\log(n),\qquad \mbox{and}\qquad \theta_{\min}\cdot\frac{(|\lambda_K|/\sqrt{\lambda_1})\|\theta\|}{K\theta_{\max}} \geq C\log(n), 
\eeq
for a large enough constant $C>0$. 
Note that the condition on $s_n$ implies
\[
\frac{\theta^2_{\min}\|\theta\|^2(|\lambda_K|/\sqrt{\lambda_1})^2}{K^8\theta^3_{\max}\|\theta\|_1}\geq C\log(n), 
\]
for a large constant $C>0$. It is straightforward that this condition guarantees \eqref{thm-main-2}. Then, the right hand side of \eqref{thm-Hamm-11} can be $o(n^{-3})$. In other words, with probability $1-o(n^{-3})$, 
\beq \label{thm-main-3}
|e_i'W\xi_k|\leq cK^{-1}\theta_i\beta_n\|\theta\|,
\eeq 
where the constant $c_1>0$ can be arbitrarily small by setting the constant $C$ in the assumption of $s_n$ to be sufficiently large. We plug it into \eqref{thm-main-4} to get, with probability $1-o(n^{-3})$, 
\[
\| H\hat{r}_i-r_i \|\leq (3c_0/16)\sqrt{K} + C_2c\sqrt{K}. 
\] 
Since $c$ can be made arbitrarily small by increasing $C$ in the assumption of $s_n$, we choose a large enough $C$ such that $C_2c<(c_0/16)\sqrt{K}$. Then, \eqref{thm-main-1} is satisfied. The claim follows directly. \qed

\bibliographystyle{plain}
\bibliography{network}
\end{document}